\documentclass[11pt]{article}
\usepackage{fancyhdr}
\usepackage{isomath}
\usepackage{amsmath}
\usepackage{amsbsy}
\usepackage{amssymb}
\usepackage{amscd}
\usepackage{amsfonts}
\usepackage{graphicx,color}
\usepackage{verbatim}
\usepackage{euscript}
\usepackage{alltt}
\usepackage{stmaryrd}

\usepackage{amsmath}
\usepackage{amsbsy}
\usepackage{amssymb}
\usepackage{amscd}
\usepackage{amsfonts}

\newcommand{\bfa}{{\mathbold a}}
\newcommand{\bfb}{{\mathbold b}}
\newcommand{\bfc}{{\mathbold c}}
\newcommand{\bfd}{{\mathbold d}}
\newcommand{\bfe}{{\mathbold e}}
\newcommand{\bff}{{\mathbold f}}

\newcommand{\bfn}{{\mathbold n}}

\newcommand{\bfs}{{\mathbold s}}

\newcommand{\bfu}{{\mathbold u}}
\newcommand{\bfv}{{\mathbold v}}

\newcommand{\bfx}{{\mathbold x}}
\newcommand{\bfy}{{\mathbold y}}
\newcommand{\bfz}{{\mathbold z}}

\newcommand{\bfA}{{\mathbold A}}
\newcommand{\bfB}{{\mathbold B}}
\newcommand{\bfC}{{\mathbold C}}
\newcommand{\bfD}{{\mathbold D}}

\newcommand{\bfF}{{\mathbold F}}
\newcommand{\bfG}{{\mathbold G}}

\newcommand{\bfI}{{\mathbold I}}
\newcommand{\bfJ}{{\mathbold J}}
\newcommand{\bfK}{{\mathbold K}}
\newcommand{\bfL}{{\mathbold L}}
\newcommand{\bfM}{{\mathbold M}}

\newcommand{\bfP}{{\mathbold P}}
\newcommand{\bfQ}{{\mathbold Q}}
\newcommand{\bfR}{{\mathbold R}}
\newcommand{\bfS}{{\mathbold S}}
\newcommand{\bfT}{{\mathbold T}}
\newcommand{\bfU}{{\mathbold U}}
\newcommand{\bfV}{{\mathbold V}}
\newcommand{\bfW}{{\mathbold W}}
\newcommand{\bfX}{{\mathbold X}}
\newcommand{\bfY}{{\mathbold Y}}
\newcommand{\bfZ}{{\mathbold Z}}

\newcommand{\vphi}{{\varphi}}
\newcommand{\eps}{{\varepsilon}}

\newcommand{\beq}{\begin{equation}}
\newcommand{\eeq}{\end{equation}}
\newcommand{\beqs}{\begin{eqnarray}}
\newcommand{\eeqs}{\end{eqnarray}}
\newcommand{\beql}{\begin{equation} \label}
\newcommand{\half}{\frac{1}{2}}

\newcommand{\calD}{{\cal D}}

\newcommand{\bfchi}{\mathbold{\chi}}

\newcommand{\bfnu}{\mathbold{\nu}}
\newcommand{\bfalpha}{\mathbold{\alpha}}
\newcommand{\bfomega}{\mathbold{\omega}}

\newcommand{\bfvphi}{\mathbold{\varphi}}

\newcommand{\bfPi}{\mathbold{\Pi}}

\newcommand{\bfLambda}{\mathbold{\Lambda}}

\newcommand{\bfOmega}{\mathbold{\Omega}}

\newcommand{\bfSigma}{\mathbold{\Sigma}}
\newcommand{\bfDelta}{\mathbold{\Delta}}
\newcommand{\bfveps}{\mathbold{\varepsilon}}

\newcommand{\grad}{\mathop{\rm grad}\nolimits}
\newcommand{\divergence}{\mathop{\rm div}\nolimits}
\newcommand{\curl}{\mathop{\rm curl}\nolimits}

\usepackage[margin=1in]{geometry}
\newcommand{\del}{\ensuremath{\partial}}
\newcommand{\hpsi}{\ensuremath{\psi}}

\newcommand{\pbPibf}{\ensuremath{{}^\star\!{\bfPi}}}
\newcommand{\pbPi}{\ensuremath{{}^\star\!{\Pi}}}

\newcommand{\delpsi}[1]{\ensuremath{\left( {\partial_{#1} \psi} \right)}}

\begin{document}
\title{Continuum mechanics of the interaction of phase boundaries and dislocations in solids
\footnote{Published in Springer Proceedings in Mathematics and Statistics, vol. 137, pp. 123 - 165, 2015; for Workshop on Differential Geometry and Continuum Mechanics held at the Intl. Centre for Mathematical Sciences in Edinburgh, 2013. Ed: G. Q Chen, M. Grinfeld, R.J. Knops.}}
\author{Amit Acharya$^1$\footnote{email: acharyaamit@cmu.edu} and Claude Fressengeas$^2$\footnote{claude.fressengeas@univ-lorraine.fr}\\\\
$^1$Carnegie Mellon University, Pittsburgh, USA\\
$^2$Laboratoire d'Etude des Microstructures et \\
de M\'ecanique des Mat\'eriaux (LEM3),\\
Universit\'e de Lorraine/CNRS, \\Ile du Saulcy, 57045 Metz
Cedex, France}

\date{January 2, 2015}
\maketitle
\begin{abstract}
\noindent The continuum mechanics of line defects representing singularities due to terminating discontinuities of the elastic displacement and its gradient field is developed. The development is intended for application to coupled phase transformation, grain boundary, and plasticity-related phenomena at the level of individual line defects and domain walls. The continuously distributed defect approach is developed as a generalization of the discrete, isolated defect case. Constitutive guidance for equilibrium response and dissipative driving forces respecting frame-indifference and non-negative mechanical dissipation is derived. A differential geometric interpretation of the defect kinematics is developed, and the relative simplicity of the actual adopted kinematics is pointed out. The kinematic structure of the theory strongly points to the incompatibility of dissipation with strict deformation compatibility.

\end{abstract}

\section{Introduction}

Whether due to material contrast or material instability, there
are many situations in solid mechanics that necessitate the
consideration of 2-d surfaces across which a distortion measure is
discontinuous. By a \emph{distortion} we refer to measures akin to
a deformation `gradient' except, in many circumstances, such a
measure is not the gradient of a vector field; we refer to a 2-d
surface  of discontinuity of a distortion measure as a \emph{phase
boundary} (which, of course, includes a grain boundary as a
special case). The more familiar situation in conventional theory
(i.e. nonlinear elasticity, rate-independent macroscopic
plasticity) is when the distortion field corresponds to the
gradient of a continuous displacement field, but one could, and
here we will, consider the presence of dislocations, or a
discontinuity in the elastic displacement field, as well when
necessary. We are particularly interested in situations where the
phase boundary discontinuity actually terminates along a curve on
the surface or, more generally, shows in-plane gradients along the
surface. We consider such terminating curves as phase boundary
\emph{tips} and the more general case as a continuously
distributed density of tips and their coupling to dislocations. We refer to the phase boundary tip
curves as \emph{generalized disclinations} (or g.disclinations; a
(classical) disclination in solids corresponds to the tip
constituting the termination of a pure rotation discontinuity).
Concrete physical situations where the kinematic construct we have
just outlined occur are commonplace. In connection to fundamental,
(un)loaded, microstructure of materials, such terminating boundaries
(or domain walls) occur as grain boundaries and triple junction
lines in polycrystalline metals
\cite{de1972partial,bieler2012role,lehman2010cyclic,hefferan2012observation} or layered
polymeric materials \cite{listak2006stabilization,ryu2012role}. As
agents of failure, some examples are weak interfaces between
matrix and fiber in fiber-reinforced polymer composites, or two
such phase boundaries spaced closely apart enclosing a matrix weak
zone in such materials, e.g. crazed inclusions and shear bands. Of
course, deformation bands (especially shear bands) are just as
commonplace in the path to failure in metallic materials and
granular materials. More mundane situations arise in understanding
stress singularities at sharp corners of inclusions in a matrix of
dissimilar material in a linear elastic context.

The conditions for the emergence of phase boundaries/localized deformation bands are by now well-understood, whether in the theory of inelastic deformation localization, e.g. \cite{hill1975bifurcation, rice1976localization, peirce1982analysis} or solid-solid phase transformations, e.g. \cite{knowles1978failure, james1981finite, abeyaratne2006evolution}. On the other hand, there does not exist a theory today to represent the kinematics and dynamics of the terminating lines of such phase boundaries and the propagation of these boundary-tips with the associated coupling to dislocations\footnote{ We note here the work of \cite{simha1998kinetics} that discusses kinetics of junctions of phase-boundaries under the assumption that the deformation is compatible while \cite{groger2010incompatibility}, \cite{paul2008non} discuss coupling of strain incompatibilities with phase-transformation without explicit kinematic consideration of strain-gradient incompatibilities due to terminating phase boundaries.}. This can be of primary importance in understanding progressive damage, e.g. onset of debonding at fiber-matrix interfaces, extension of shear bands or crazes, or the stress concentrations produced at five-fold twin junctions, or grain boundary triple lines. It is the goal of this paper to work out the general continuum mechanics of coupled phase boundary and slip (i.e. regularized displacement-gradient and displacement discontinuities), taking into account their line defects which are g.disclinations and dislocations. The developed model is expected to be of both theoretical and practical use in the study of the coupling of the structure and motion of phase boundaries coupled to dislocation and kink-like defects e.g. \cite{hirth2011compatibility, wang2013defective, simon2010multiplication}.

A corresponding `small deformation' theory has been worked out in \cite{acharya2012coupled}. It was not clear to us then whether one requires a theory with couple stress or not and both thermodynamically admissible possibilities were outlined there. We now believe that dealing with g.disclinations requires mechanics mediated by torque balance
\footnote{However, a dislocation-only defect model does not require any consideration of torque balance or couple stresses, as shown in \cite{acharya2011microcanonical, acharya2012coupled} and in Sec \ref{sec_rev_diss}.}
and, therefore, in this paper, we only consider models where couple stresses also appear. A dissipative extension of disclination-dislocation theory due to deWit \cite{de1970linear} has been developed in \cite{fressengeas2011elasto, upadhyay2011grain, upadhyay2013elastic} as well as the first numerical implementations for the theory with application to understanding grain-boundary mechanics \cite{taupin2012grain, taupin_shear_coupled}. While we focus on continuously distributed defect densities, it is to be understood that we include in our setting the modeling of individual defect lines as non-singular localizations of these density fields along space curves.

The concept of classical disclinations and dislocations arose in the work of Weingarten and Volterra (cf. \cite{nabarro1967theory}) from the specific question of characterizing the displacement and rotation jumps across a surface of a multiply connected region with a hole, when the displacement field is required to be consistent with a prescribed twice differentiable strain (metric field) field; a well-developed static theory exists \cite{romanov2009application} as well as a very sophisticated topological theory, full of subtle but difficult insights, due to Klem\'an and Friedel \cite{kleman2008disclinations}. While self-contained in itself, this question does not suffice for our purposes in understanding phase boundaries, since these can, and often necessarily, involve jumps in the strain field. Nevertheless, the differential geometry of coupled dislocations and so-called disclinations have been the subject of extensive enquiry, e.g. \cite{kondo1955non, bilby1960continuous, kroner1992gauge, clayton2006modeling}, and therefore we show how our g.disclinations can be placed in a similar differential geometric context, while pointing out the main differences from the standard treatment. The differences arise primarily from a desire to achieve relative simplicity by capitalizing on the available Euclidean structure of the ambient space in which we do our mechanics directed towards applications. 

The remainder of the paper is organized as follows. In Section 2 we provide a list of notation. In Section 3 we develop a fundamental kinematic decomposition relevant for our work. In Section 4 we develop the governing mechanical equations. In Section 5 we examine consequences of material frame-indifference (used synonymously with invariance under superposed rigid body motions) and a dissipation inequality for the theory, ingredients of which provide a critical check on the finite deformation kinematics of the proposed evolution equations for defect densities. Section 6 describes a small deformation version of the model. In Section 7 we provide a differential geometric interpretation of our work. Some concluding observations are recorded in Section 8.

\begin{figure}
\begin{center}
\includegraphics[height=0.3\textheight,width=0.9\textwidth,angle=0]{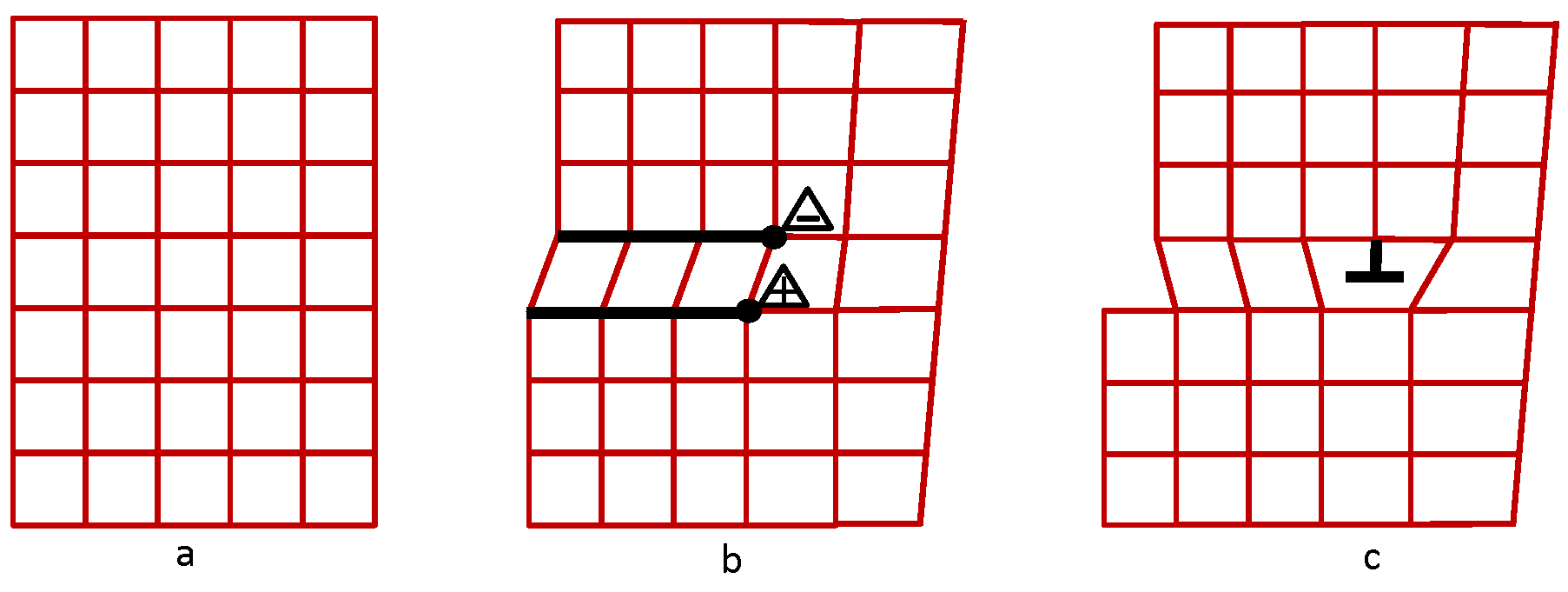}
\caption{Path to an idealized edge dislocation nucleation (c)
involving a deformation discontinuity, achieved through the
formation of a g.disclination dipole (b) in a continuous
deformation with two surfaces of strain discontinuity of an unstretched atomic configuration (a). Here, a
continuous deformation (b) of the original configuration (a)
refers to the preservation of all nearest neighbors signified by
bond connections; a discontinuous deformation (c) refers to a change in topology
of bond connections.} \label{g.disclin_dipole_disloc}
\end{center}
\end{figure}

Finally, in order to provide some physical intuition for the new
kinematic objects we have introduced before launching into their
continuum mechanics, we demonstrate (Figure
\ref{g.disclin_dipole_disloc}) a possible path to the nucleation
of an edge dislocation in a lattice via the formation of a
g.disclination dipole. It is then not surprising that point-wise
loss of ellipticity criteria applied to continuum response
generated from interatomic potentials  can bear some connection to
predicting the onset of dislocation nucleation
\cite{li2002atomistic, zhu2004predictive}. 


\section{Notation}\label{notation}

A superposed dot on a symbol represents a material time
derivative. The statement $a := b$ indicates that $a$ is defined
to be equal to $b$. The summation convention is implied unless
otherwise mentioned. We denote by $\bfA \bfb$ the action of the
second-order (third-order, fourth-order) tensor $\bfA$ on the
vector (second-order tensor, second-order tensor) $\bfb$,
producing a vector (vector, second-order tensor). A $\cdot$
represents the inner product of two vectors, a $:$ represents the
trace inner product of two second-order tensors (in rectangular
Cartesian components, $\bfA : \bfD = A_{ij} D_{ij}$) and matrices
and the contraction of the last two indices of a third-order
tensor with a second order tensor. The symbol $\bfA\bfD$
represents tensor multiplication of the second-order tensors
$\bfA$ and $\bfD$. The notation $(\cdot)_{\mbox{\tiny sym}}$ and
$(\cdot)_{\mbox{\tiny skw}}$ represent the symmetric and skew
symmetric parts, respectively, of the second order tensor
$(\cdot)$. We primarily think of a third-order tensor as a linear
transformation on vectors to the space of second-order tensors. A
\emph{transpose} of a third-order tensor is thought of as a linear
transformation on the space of second order tensors delivering a
vector and defined by the following rule: for a third-order tensor
$\bfB$
\[
\left( \bfB^T \bfD \right) \cdot \bfc = \left( \bfB \bfc \right) : \bfD,
\]
for all second-order tensors $\bfD$ and vectors $\bfc$.

The symbol $\divergence$ represents the divergence, $\grad$ the
gradient, and $\divergence \grad$ the Laplacian on the current
configuration. The same words beginning with a Latin uppercase
letter represent the identical derivative operators on a reference
configuration. The $\curl$ operation and the cross product of a
second-order tensor and a vector are defined in analogy with the
vectorial case and the divergence of a second-order tensor: for a
second-order tensor $\bfA$, a third-order tensor $\bfB$, a vector
$\bfv$, and spatially constant vector fields $\bfb$, $\bfc$, and a
spatially uniform second-order tensor field $\bfD$,
\begin{equation*}
    \begin{split}
        \bfc \cdot \left( \bfA \times \bfv \right) \bfb  &= \left[ \left( \bfA^{T} \bfc \right) \times \bfv \right] \cdot \bfb \quad \forall \bfb, \bfc \\
            \bfD    : \left( \bfB \times \bfv \right) \bfb &= \left[ \left( \bfB^T \bfD \right) \times \bfv \right] \cdot \bfb \quad \forall \bfD, \bfb
\\
        \left(\divergence\bfA\right) \cdot \bfc &= \divergence \left( \bfA^{T} \bfc \right) \quad \forall \bfc  \\
                \left(\divergence\bfB\right) : \bfD &= \divergence \left( \bfB^{T} \bfD \right) \quad \forall \bfD
\\
        \bfc \cdot \left(\curl\bfA\right)\bfb &= \left[ \curl\left(\bfA^{T}\bfc\right) \right] \cdot \bfb \quad \forall \bfb, \bfc
\\
\bfD : \left(\curl\bfB\right)\bfb &= \left[ \curl\left(\bfB^{T}\bfD\right) \right] \cdot \bfb \quad \forall \bfb, \bfD
    \end{split}
\end{equation*}
In rectangular Cartesian components,
\begin{equation*}
 \begin{split}
    \left(\bfA\times\bfv\right)_{im} &= e_{mjk} A_{ij} v_{k}    \\
    \left(\bfB\times\bfv\right)_{irm} &= e_{mjk} B_{irj} v_{k}  \\
    \left(\divergence\bfA\right)_{i} &= A_{ij,j}                \\
    \left(\divergence\bfB\right)_{ij} &= B_{ijk,k}              \\
    \left(\curl\bfA\right)_{im} &= e_{mjk} A_{ik,j}             \\
    \left(\curl\bfB\right)_{irm} &= e_{mjk} B_{irk,j}               \\
 \end{split}
\end{equation*}
where $e_{mjk}$ is a component of the third-order alternating tensor $\bfX$.
Also, the vector $\bfX\bfA\bfD$ is defined as
\begin{equation*}
    \left(\bfX\bfA\bfD\right)_{i} = e_{ijk} A_{jr} D_{rk}.
\end{equation*}
The spatial derivative for the component representation is with
respect to rectangular Cartesian coordinates on the current
configuration of the body. Rectangular Cartesian coordinates on
the reference configuration will be denoted by uppercase Latin
indices. For manipulations with components, we shall always use
such rectangular Cartesian coordinates, unless mentioned otherwise. Positions of particles are
measured from the origin of this arbitrarily fixed Cartesian
coordinate system.

For a second-order tensor $\bfW$, a third-order tensor $\bfS$ and
an orthonormal basis $\{ \bfe_i, i = 1,2,3 \}$ we often use the
notation
\[
\left( \bfW \bfS^{2T} \right) = W_{lp}S_{rpk} \bfe_r \otimes \bfe_l \otimes \bfe_k \ \ ; \ \ \left( \bfW \bfS^{2T} \right)_{rlk}  :=  W_{lp}S_{rpk}.
\]


The following list describes some of the mathematical symbols we use in this paper.
\\
\noindent $\bfx$: current position\\
$\bfF^e$: elastic distortion tensor ($2^{nd}$-order)\\
$\bfW = \left(\bfF^e\right)^{-1}$: inverse of elastic 1-distortion tensor ($2^{nd}$-order)\\
$\bfS$: eigenwall tensor ($3^{rd}$-order)\\
$\bfY$: inverse-elastic 2-distortion tensor ($3^{rd}$-order)\\
$\bfalpha$: dislocation density tensor ($2^{nd}$-order)\\
$\bfPi$: g.disclination density tensor ($3^{rd}$-order)\\
$\bfv$: material velocity\\
$\bfL$: velocity gradient\\
$\bfD=\bfL_{sym}$: rate of deformation tensor\\
$\bfOmega= \bfL_{skw}$: spin tensor\\
$\bfomega = - \frac{1}{2} \bfX:\bfOmega = \frac{1}{2}\, curl \,\bfv$: half of the vorticity vector\\
$\bfM= grad \, \bfomega$: vorticity gradient tensor\\
$\bfJ = grad \,\bfW$: gradient of i-elastic distortion\\
$\bfV^{\Pi}$: g.disclination velocity\\
$\bfV^{\alpha}$: dislocation velocity\\
$\bfV^{S}$: eigenwall velocity\\
$\bfT$: Cauchy stress tensor\\
$\bfLambda$: couple stress tensor\\
$\bfK$: external body moment per unit mass\\
$\bfb$: external body force per unit mass\\
$\rho$: mass density\\
$\psi$: free energy per unit mass
\section{Motivation for a fundamental kinematic decomposition}\label{motiv}
\begin{figure}
\begin{center}
\includegraphics[height=0.3\textheight,width=0.9\textwidth,angle=0]{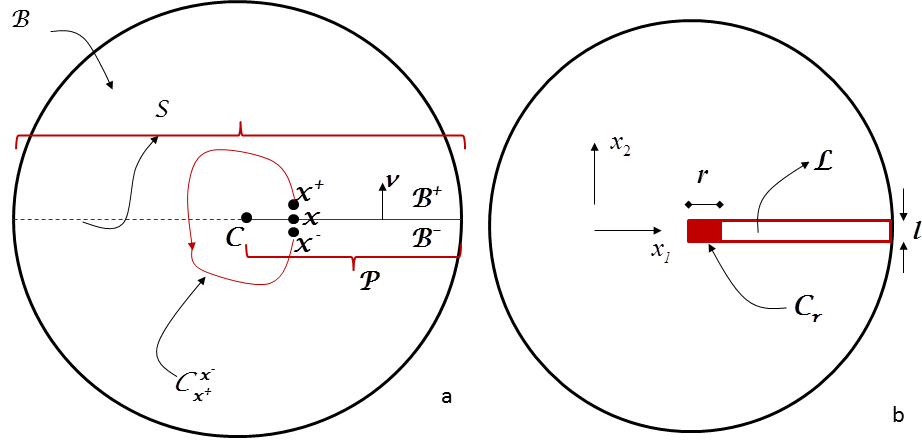}
\caption{Classical terminating discontinuity and its physical regularization.}
\label{classical_reg_discont}
\end{center}
\end{figure}

With reference to Figure \ref{classical_reg_discont}a representing a cross-section of a body, suppose we
are given a tensor field $\bfvphi$ ($0^{th}$-order and up) that
can be measured unambiguously, or computed from measurements
without further information, at most points of a domain $\cal{B}$.
Assume that the field $\bfvphi$ is smooth everywhere except having
a terminated discontinuity of constant magnitude across the
surface $\cal{S}$. Denote the terminating curve of the
discontinuity on the surface $\cal{S}$ as $\cal{C}$. We think of
the subset $\cal{P}$ of $\cal{S}$ across which a non-zero jump
exists as a \emph{wall} of the field $\bfvphi$ and the curve
$\cal{C}$ as a \emph{line defect} of the field $\bfvphi$. Physical
examples of walls are domain walls, grain boundaries, phase
boundaries, slip boundaries and stacking faults (surfaces of
displacement discontinuity); those of defect lines are vortices,
disclinations, g.disclinations, and dislocations. 
Let $\bfnu$ be a unit normal field on $\cal{S}$, with arbitrarily
chosen orientation. Let $\cal{B}^+$ be the subset of $\cal{B}$
into which $\bfnu$ points; similarly, let $\cal{B}^-$ be the
subset of $\cal{B}$ into which $-\bfnu$ points. Let $\bfx$ be a
point on $\cal{P}$. Let $\bfx^+ \in \cal{B}^+$ and $\bfx^- \in
\cal{B}^-$ be points arbitrarily close to $\bfx$ but not $\bfx$,
and let $\bfvphi(\bfx^+) = \bfvphi^+$ and $\bfvphi(\bfx^-) =
\bfvphi^-$. Join $\bfx^+$ to $\bfx^-$ by any contour
$C_{\bfx^+}^{\bfx^-}$ encircling $\cal{C}$. Then
\begin{equation}\label{topol}
\int_{C_{\bfx^+}^{\bfx^-}} grad \  \bfvphi \cdot d\bfx = \bfvphi^- - \bfvphi^+ =: - \llbracket \bfvphi \rrbracket.
\end{equation}
Note that by hypothesis $\llbracket \bfvphi \rrbracket$ is
constant on $\cal{P}$ so that regardless of how close $\bfx$ is to
$\cal{C}$, and how small the non-zero radius of a circular contour
$C_{\bfx^+}^{\bfx^-}$ is, the contour integral takes the same
value. This implies that $ |grad \ \bfvphi(\bfy)| \rightarrow
\infty$ as $\bfy \rightarrow \cal{C}$ with $\bfy \in \cal{B}
\backslash \cal{C}$ 
\footnote{As an aside, this observation also shows why the typical
assumptions made in deriving transport relations for various types
of control volumes containing a shock surface do not hold when the
discontinuity in question is of the `terminating jump' type being
considered here.}. 
Our goal now is to define a field $\bfA$ that is a physically
regularized analog of $grad \ \bfvphi$; we require $\bfA$ to not
have a singularity but possess the essential topological property
(\ref{topol}) if $grad \, \bfvphi$ were to be replaced there with
$\bfA$. For instance, this would be the task at hand if, as will
be the case here, $\bfA$ is an ingredient of a theory and initial
data for the field needs to be prescribed based on available
observations on the field $\bfvphi$, the latter as described above.

It is a physically natural idea to regularize the discontinuity on
$\cal{P}$ by a field on $\cal{B}$ that has support only on a thin
layer around $\cal{P}$. We define such a field as follows (Figure
\ref{classical_reg_discont}b). For simplicity, assume all fields
to be uniform in the $x_3$-direction. Let the layer $\cal{L}$ be
the set of points
\[
{\cal{L}} = \left\{\bfy \in {\cal{B}} : \bfy = \bfx + h\, \bfnu (\bfx), -l/2 \leq h \leq l/2, \bfx \in \cal{P} \right\}.
\]
Let the $x_1$ coordinate of $\cal{C}$ be $x^0$. Define the \emph{strip} field
\[
\bfW\bfV(\bfx) = \begin{cases} f(x_1) \frac{\left\{\bfvphi^-(x_1) - \bfvphi^+(x_1)\right\}}{l} \otimes \bfnu(x_1), & \mbox{if } \bfx \in {\cal{L}} \\
\bf0, & \mbox{if } \bfx \in \cal{B}\backslash\cal{L} \end{cases} \footnote{$\bfW\bfV$ is to be interpreted as the name for a single field.}
\]
where $\bfnu(x_1) = \bfe_2$ here, and
\[
f(x_1) = \begin{cases} \frac{x_1 - x^0}{r}, & \mbox{if } x^0 < x_1 \leq x^0 + r \\
1, & \mbox{if } x^0+r \leq x_1. \end{cases}
\]
In the above, the layer width $l$ and the defect-core width $r$ are considered as given physical parameters. We now define $\bfA$ as
\begin{equation}\label{funda}
\bfA := grad\, \bfB + \bfW\bfV \ \ \mbox{in } {\cal{B}},
\end{equation}
where $\bfB$ is at least a continuous and piecewise-smooth
potential field in $\cal{B}$, to be determined from further
constraints within a theoretical structure as, for example, we
shall propose in this paper.

Let $n$ be the order of the tensor field $\bfvphi$. A small
calculation shows that the only non-vanishing component(s) of
$curl\, \bfW\bfV$ is
\[
(curl\, WV)_{i_1 \cdots i_n 3} = e_{312} \frac{\del f}{\del x_1} \left[ \frac{- \llbracket \vphi \rrbracket_{i_1 \cdots i_n}}{l} \right] = WV_{i_1 \cdots i_n 2,1}\ \footnote
{Here it is understood that if $n = 0$ then the symbol $i_1
\cdots i_n$ correspond to the absence of any indices and the $curl$ of the higher-order tensor field is understood as the natural analog of the second-order case defined in Section \ref{notation}.}
\]
and this is
non-zero only in the \emph{core cylinder} defined by
\[
{\cal{C}}_r = \left\{ \bfx : x^0 \leq x_1 \leq x^0 + r, -l/2 \leq x_2 \leq l/2 \right\}.
\]
Moreover, since $\frac{\del f}{\del x_1} = \frac{1}{r}$ in ${\cal{C}}_r$ and zero otherwise, we have
\[
\int_C \bfA \cdot d\bfx = \int_A curl \, \bfW\bfV \cdot \bfe_3 \,da = - \frac{\llbracket \bfvphi \rrbracket}{(l\cdot r)}( {l\cdot r}) = - \llbracket \bfvphi \rrbracket,
\]
for any \emph{closed} curve $C$ encircling ${\cal{C}}_r$ and $A$ is any surface patch with boundary curve $C$.

Without commitment to a particular theory with constitutive
assumptions, it is difficult to characterize further specific
properties of the definition (\ref{funda}). However, it is
important to avail of the following general intuition regarding
it. Line defects are observed in the absence of applied loads.
Typically, we are thinking of $grad \, \bfvphi$ as an elastic
distortion measure that generates elastic energy, stresses,
couple-stresses etc. Due to the fact that in the presence of line
defects as described, $grad \, \bfvphi $ has non-vanishing content
away from $\cal{P}$ in the absence of loads, if $\bfA$ is to serve
as an analogous non-singular measure, it must have a similar
property of producing \emph{residual elastic distortion} for any choice of a $grad \, \bfB$ field for a given $\bfW\bfV$ field
that contains a line defect (i.e. a non-empty subset
${\cal{C}}_r$). These possibilities can arise, for instance, from
a hypothesis on minimizing energy or balancing forces or moments.
That such a property is in-built into the definition (\ref{funda})
can be simply understood by realizing that $\bfW\bfV$ is not a
gradient and therefore cannot be entirely annihilated by $grad \,
\bfB$. To characterize this a bit further, one could invoke a
Stokes-Helmholtz type decomposition of the (localized-in-layer)
$\bfW\bfV$ field to obtain
\[
\bfW\bfV = grad \, \bfZ + \bfP \ \  \mbox{on } {\cal{B}}
\]
\[
div \, grad \, \bfZ = div \,\bfW\bfV \ \ \mbox{on } {\cal{B}}
\]
\[
grad \, \bfZ \bfn = \bfW\bfV \bfn \ \ \mbox{on } \del{\cal{B}}
\]
\begin{equation}\label{SH_decomp}
curl\, \bfP = curl \, \bfW\bfV \ \ \mbox{on } {\cal{B}}
\end{equation}
\[
div\, \bfP = \bf0 \ \ \mbox{on } {\cal{B}}
\]
\[
\bfP \bfn = \bf0 \ \ \mbox{on } \del{\cal{B}},
\]
noting the interesting fact that $grad\,\bfZ = - \bfP$ in
$\cal{B}\backslash \cal{L}$ because of the localized nature of
$\bfW\bfV$. Thus, $grad\,\bfB$ can at most negate the $grad \,
\bfZ$ part of $\bfW\bfV$ and what remains is at least a
\emph{non-localized} field $\bfP$ representing some, or in some
specific cases (e.g. screw dislocation in isotropic linear
elasticity or Neo-Hookean elasticity, \cite{acharya2001model})
all, of the off-${\cal{C}}_r$ content of the original
$grad\,\bfvphi$ field. Of course, it must be understood that the
primary advantage, within our interpretation, of utilizing $\bfA$
in place of $grad \, \bfvphi$ is that the former is non-singular,
but with the desired properties\footnote{
It is to be noted that the decomposition (\ref{SH_decomp}) is merely \emph{a} means to understand the definitions (\ref{funda}), (\ref{Y-decomp}), the latter being fundamental to the theory.}.

It should be clear now that a field with many defect lines can as
well be represented by a construct like (\ref{funda}) through
superposition of their `corresponding $\bfW\bfV$ fields',
including dipolar defect-line structures where the layer $\cal{L}$
has two-sided terminations within the body, without running all
the way to the boundary.

As a common example we may think of classical small deformation
plasticity where the plastic distortion field $\bfU^p$ may be
interpreted as $-\bfW\bfV$, the displacement field $\bfu$ as the
potential $\bfB$ and $\bfA$ as the elastic distortion $\bfU^e$. In
classical plasticity theory, the decomposition $\bfU^e = grad\,
\bfu - \bfU^p$ is introduced as a hypothesis based on
phenomenology related to 1-d stress strain curves and the notion
of permanent deformation produced in such a set-up. Our analysis
may be construed as a fundamental kinematical and microstructural
justification of such a hypothesis, whether in the presence of a
single or many, many dislocations. At finite deformations, there
is a similar decomposition for the i-elastic 1 distortion
$\bfF^{e-1} = \bfW = \bfchi + grad\,\bff$
\cite{acharya2004constitutive, acharya2011microcanonical},  where
the spatial derivative is on the current configuration and we
identify $\bfA$ with $\bfW$, $\bfZ + \bfB$ with $\bff$, and $\bfP$
with $\bfchi$.

Based on the above motivation, for the theory that follows, we
shall apply the definition (\ref{funda}) to the i-elastic
2-distortion $\bfY$ to write
\begin{equation}\label{Y-decomp}
\bfY = grad\, \bfW + \bfS,
\end{equation}
where $\bfW$ is the i-elastic 1-distortion and we refer to $\bfS$ ($3^{rd}$-order tensor) as the \emph{eigenwall} field.

What we have achieved above is a generalization of the
\emph{eigenstrain} concept of Kr\"{o}ner, Mura, and deWit. With
the gained understanding, it becomes \emph{the} natural modeling
tool for dealing with the dynamics of discontinuities and
line-singularities of first and higher-order deformation gradients
with smooth (everywhere) fields within material and geometrically
linear and nonlinear theories. The main utility of $\bfW\bfV$
fields, as will be evident later, is in providing a tool for
stating kinematically natural evolution equations for defect
densities; while they also provide regularization of nasty
singularities, such a smoothing effect can, at least in principle,
also be obtained by demanding that the jump $\llbracket \bfvphi
\rrbracket$ rise to a constant value from $0$ over a short
distance in $\cal{P}$, without introducing any new fields.

\section{Mechanical structure and dissipation}  \label{mech_structure}

\subsection{Physical notions}

This subsection has been excerpted from \cite{acharya_zhang} for the sake of completeness.

The physical model we have in mind for the evolution of the body
is as follows. The body consists of a fixed set of atoms. At any
given time each atom occupies a well defined region of space and
the collection of these regions (at that time) is
well-approximated by a connected region of space called a
configuration. We \emph{assume} that any two of these
configurations can necessarily be connected to each other by a
continuous mapping. The temporal sequence of configurations
occupied by the set of atoms are further considered as
parametrized by increasing time to yield a \emph{motion} of the
body. A fundamental assumption in what follows is that the mass
and momentum of the set of atoms constituting the body are
transported in space by this continuous motion. For simplicity, we
think of each spatial point of the configuration corresponding to
the body in the as-received state for any particular analysis as a
set of `material particles,' a particle generically denoted by
$\bfX$.

Another fundamental assumption related to the motion of the atomic
substructure is as follows. Take a  spatial point $\bfx$ of a
configuration at a given time $t$. Take a collection of atoms
around that point in a spatial volume of fixed extent, the latter
independent of $\bfx$ and with size related to the spatial scale
of resolution of the model we have in mind. Denote this region as
${\cal{D}}_c(\bfx, t)$; this represents the `box' around the base
point $\bfx$ at time $t$. We now think of relaxing the set of
atoms in ${\cal{D}}_c(\bfx, t)$ from the constraints placed on it
by the rest of the atoms of the whole body, the latter possibly
externally loaded. This may be achieved, in principle at least, by
removing the rest of the atoms of the body or, in other words, by
ignoring the forces exerted by them on the collection within
${\cal{D}}_c(\bfx, t)$. This (thought) procedure generates a
unique placement of the atoms in ${\cal{D}}_c(\bfx, t)$ denoted by
\textsf{A}$_{\bfx}$ with no forces in each of the atomic bonds in
the collection.

We now imagine immersing \textsf{A}$_{\bfx}$ in a larger
collection of atoms (without superimposing any rigid body
rotation), ensuring that the entire collection is in a zero-energy
ground state (this may require the larger collection to be `large
enough' but not space-filling, as in the case of amorphous
materials (cf. \cite{kleman1979tentative}). Let us assume that as
$\bfx$ varies over the entire body, these larger collections, one
for each $\bfx$, can be made to contain identical numbers of
atoms. Within the larger collection corresponding to the point
$\bfx$, let the region of space occupied by \textsf{A}$_{\bfx}$ be
approximated by a connected domain ${\cal{D}}_r^{pre}(\bfx, t)$,
containing the same number of atoms as in ${\cal{D}}_c(\bfx, t)$.
The spatial configuration ${\cal{D}}_r^{pre}(\bfx, t)$ may
correctly be thought of as stress-free. Clearly, a deformation can
be defined mapping the set of points ${\cal{D}}_c(\bfx, t)$ to
${\cal{D}}_r^{pre}(\bfx, t)$. We now \emph{assume} that this
deformation is well approximated by a homogeneous deformation.

Next, we assume that the set of these larger collections of
relaxed atoms, one collection corresponding to each $\bfx$ of the
body,  differ from each other only in orientation, if
distinguishable at all. We choose one such larger collection
arbitrarily, say \textsf{C}, and impose the required rigid body
rotation to each of the other collections to orient them
identically to \textsf{C}. Let the obtained configuration after
the rigid rotation of ${\cal{D}}_r^{pre}(\bfx, t)$ be denoted by
${\cal{D}}_r(\bfx, t)$.

We denote the  \emph{gradient of the homogeneous deformation
mapping ${\cal{D}}_c(\bfx, t)$ to ${\cal{D}}_r(\bfx, t)$} by
$\bfW(\bfx, t)$, the i-elastic 1-distortion at $\bfx$ at time $t$.

What we have described above is an embellished version of the
standard fashion of thinking about the problem of defining elastic
distortion in the classical theory of finite elastoplasticity
\cite{lee}, with an emphasis on making a connection between the
continuum mechanical ideas and discrete atomistic ideas as well as
emphasizing that no ambiguities related to spatially inhomogeneous
rotations need be involved in defining the field $\bfW$
\footnote{Note that the choice of \textsf{C} affects  the $\bfW$
field at most by a superposed \emph{spatio-temporally uniform}
rotation field.}. 
However, our physical construct requires no choice of a reference
configuration or a `multiplicative decomposition' of it into
elastic and plastic parts to be invoked
\cite{acharya2004constitutive}. In fact, \emph{there is no notion
of a plastic deformation $\bfF^p$ invoked in our model}. Instead,
as we show in Section \ref{disloc_density_evol}
(\ref{add_decomp}), an additive decomposition of the velocity
gradient into elastic and plastic parts emerges naturally in this
model from the kinematics of dislocation motion representing
conservation of Burgers vector content in the body.

Clearly, the field $\bfW$ need not be a gradient of a vector field
at any time. Thinking of this ielastic 1-distortion field $\bfW$
on the current configuration at any given time as the $\bfvphi$
field of Section \ref{motiv}, the i-elastic 2-distortion field
$\bfY$ is then defined as described therein.

It is important to note that if a material particle $\bfX$ is
tracked by an individual trajectory $\bfx(t)$ in the motion (with
$\bfx(0) = \bfX$), the family of neighborhoods
${\cal{D}}_c(\bfx(t), t)$ parametrized by $t$ in general can
contain vastly different sets of atoms compared to the set
contained initially in ${\cal{D}}_c(\bfx(0), 0)$. The intuitive
idea is that the connectivity, or nearest neighbor identities, of
the atoms that persist in ${\cal{D}}_c(\bfx(t), t)$ over time
remains fixed only in purely elastic motions of the body.

\subsection{The standard continuum balance laws}
For any fixed set of material particles occupying the volume
$B(t)$ at time $t$ with boundary $\partial B (t)$ having outward
unit normal field $\bfn$
\begin{equation*}
    \begin{split}
    \dot{\overline{\int_{B(t)} \rho \ dv}} &= 0, \\
    \dot{\overline{\int_{B(t)} \rho \bfv \ dv}} &= \int_{\partial B(t)} \bfT \bfn \ da + \int_{B(t)} \rho \bfb \ dv, \\
    \dot{\overline{\int_{B(t)} \rho \left( \bfx \times \bfv  \right) \, dv}} &= \int_{\partial B(t)} \left(\bfx \times \bfT + \bfLambda \right) \bfn \ da + \int_{B(t)} \rho \left( \bfx \times \bfb + \bfK \right) \ dv,
    \end{split}
\end{equation*}
represent the statements of balance of mass, linear and angular
momentum, respectively. We re-emphasize that it is an assumption
that the actual mass and momentum transport of the underlying
atomic motion can be adequately represented through the material
velocity and density fields governed by the above statements (with
some liberty in choosing the stress and couple-stress tensors).
For instance, in the case of modeling fracture, some of these
assumptions may well require revision.

Using Reynolds' transport theorem, the corresponding local forms for these equations are:
\begin{equation}\label{balance_laws}
    \begin{split}
        \dot{\rho} + \rho \divergence \bfv & = 0\\
        \rho\dot{\bfv} &= \divergence \bfT + \rho\bfb \\
        \bf0 & = \divergence \bfLambda - \bfX:\bfT + \rho\bfK.
    \end{split}
\end{equation}
Following \cite{mindlin1962effects}, the external power supplied to the body at any given time is expressed as:
\begin{equation*}
    \begin{split}
    P(t) &= \int_{B(t)} \rho \bfb \cdot \bfv \, dv + \int_{\partial B(t)} \left(\bfT\bfn\right)\cdot\bfv \, da  + \int_{\partial B(t)} \left(\bfLambda\bfn\right)\cdot\bfomega \, da +  \int_{B(t)} \rho \bfK \cdot \bfomega \, dv\\
      &= \int_{B(t)} \left( \rho\bfv\cdot\dot{\bfv}\right)\ dv +  \int_{B(t)} \left( \bfT : \bfD \ + \  \bfLambda : \bfM  \right) \, dv,
    \end{split}
\end{equation*}
where Balance of linear momentum and angular momentum have been used. On defining the kinetic energy and the free energy of the body as
\begin{equation*}
    \begin{split}
        K &= \int_{B(t)} \half \left( \rho \bfv\cdot\bfv \right) \ dv,\\
        F & =\int_{B(t)} \rho \psi \ dv,
    \end{split}
\end{equation*}
respectively, and using Reynolds' transport theorem, we obtain the mechanical dissipation
\begin{equation}
    \label{eqn:dissipation}
    \textsf{D} := P - \dot{\overline{K+F}} = \int_{B(t)} \left( \bfT:\bfD + \bfLambda: \bfM - \rho\dot{\psi} \right) \ dv.
\end{equation}
The first equality above shows the distribution of applied mechanical power into kinetic, stored and dissipated parts.
The second equality, as we show subsequently, is used to provide guidance on constitutive structure.

\subsection{G.disclination density and eigenwall evolution}\label{g_disc_ewall}

The natural measure of g.disclination density is
\[
curl \, \left( \bfY - grad \, \bfW \right) = curl \, \bfS = \bfPi.
\]
It characterizes the closure failure of integrating $\bfY$ on closed contours in the body:
\[
\int_a \bfPi \bfn da = \int_c \bfY \, d\bfx,
\]
where $a$ is any area patch with closed boundary contour $c$ in
the body. Physically, it is to be interpreted as a density of
lines (threading areas) in the current configuration, carrying a
tensorial attribute that reflects a jump in $\bfW$. As such, it is
reasonable to postulate, before commitment to constitutive
equations, a tautological evolution statement of balance for it in
the form of ``rate of change = what comes in - what goes out +
what is generated." Since we are interested in nonlinear theory
consistent with frame-indifference and non-negative dissipation,
it is more convenient to work with the measure
\begin{eqnarray}\label{Pi_star_def}
\pbPibf & := & curl \left( \bfW \bfS^{2T} \right)\\\nonumber
\left( \bfW \bfS^{2T} \right)_{rlk} & := & W_{lp}S_{rpk} \\\nonumber
\pbPi_{rli} & = & e_{ijk} \left[ W_{lp} S_{rpk} \right]_{,j} = e_{ijk} \left[ W_{lp} \left( Y_{rpk} - W_{rp,k} \right) \right]_{,j},
\end{eqnarray}
(cf. \cite{acharya-dayal}), and follow the arguments in
\cite{acharya2011microcanonical} to consider a conservation
statement for a \emph{density of lines} of the form
\begin{equation}\label{Pi_int_form}
\dot{\overline{ \int_{a\left(t \right)} \pbPibf \bfn \, da }}  =  - \int_{c(t)} \bfPi \times \bfV^{\Pi} \, d \bfx.
\end{equation}
Here, $a(t)$ is the area-patch occupied by an arbitrarily fixed
set of material particles at time $t$ and $c(t)$ is its closed
bounding curve and the statement is required to hold for all such
patches. $\bfV^{\Pi}$ is the \emph{g.disclination velocity} field,
physically to be understood as responsible for transporting the
g.disclination line density field in the body.

Arbitrarily fix an instant of time, say $s$, in the motion of a
body and let $\bfF_s$ denote the time-dependent deformation
gradient field corresponding to this motion with respect to the
configuration at the time $s$. Denote positions on the
configuration at time $s$ as $\bfx_s$ and the image of the area
patch $a(t)$ as $a(s)$. We similarly associate the closed curves
$c(t)$ and $c(s)$. Then, using the definition (\ref{Pi_star_def}),
(\ref{Pi_int_form}) can be written as
\begin{eqnarray}\nonumber
\dot{\overline{ \int_{a\left(t \right)} \pbPibf \bfn \, da }} & + & \int_{c(t)} \bfPi \times \bfV^{\Pi} \, d \bfx =  \dot{ \overline{ \int_{c(t)} \bfW \bfS^{2T}  \, d\bfx  }} + \int_{c(t)} \bfPi \times \bfV^{\Pi} \, d \bfx \\\nonumber
  & = & \int_{c(s)} \left[ \dot{ \overline{ \bfW \bfS^{2T} \bfF_s }} + \left( \bfPi \times \bfV^{\Pi} \right)\bfF_s \right] d\bfx_s\\\nonumber
  & = & \int_{c(t)} \left[ \dot{ \overline{ \bfW \bfS^{2T} \bfF_s }}\bfF^{-1}_{s} + \bfPi \times \bfV^{\Pi} \right] d\bfx = \bf0
\end{eqnarray}
which implies
\[
\dot{ \overline{ \bfW \bfS^{2T} \bfF_s }}\bfF^{-1}_{s} = - \bfPi \times \bfV^{\Pi} + grad \, \bfSigma,
\]
where $\bfSigma$ is an arbitrary second-order tensor field with
physical dimensions of strain rate (i.e. $1/Time$) that we will
subsequently specify to represent grain/phase boundary motion
transverse to itself. Finally, choosing $s = t$, we arrive at the
following local evolution equation for $\bfS$:
\[
\stackrel{\diamond}{\bfS} := \dot{\bfW} \bfS^{2T} + \bfW \dot{\bfS^{2T}} + \bfW \bfS^{2T} \bfL = - \bfPi \times \bfV^{\Pi} + grad \, \bfSigma.
\]
The local form of (\ref{Pi_int_form}) is
\begin{equation}\label{Pi_loc}
\stackrel{\circ}{\overline{\pbPibf}} := \left( div\, \bfv \right) \pbPibf + \dot{\overline{\pbPibf}} - \pbPibf \bfL^T = -curl \, \left( \bfPi \times \bfV^{\Pi} \right).\footnote
{An important feature of conservation statements for signed
`topological charge' as here is that even without explicit source
terms nucleation (of loops) is allowed. This fact, along with the
coupling of $\bfPi$ to the material velocity field through the convected derivative provides an
avenue for predicting homogeneous nucleation of line defects. In
the dislocation-only theory, some success has been achieved with
this idea in ongoing work.} 
\end{equation}

Finally, we choose $\bfSigma$ to be
\[
\bfSigma := \bfW \bfS^{2T} \bfV^S \ \ ;\ \ \Sigma_{ij} = W_{jr}S_{irk}V^S_{k},
\]
where $\bfV^S$ is the \emph{eigenwall velocity} field that is
physically to be interpreted as transporting the eigenwall field
$\bfS$ transverse to itself. This may be heuristically justified
as follows: the eigenwall field represents a gradient of i-elastic
distortion in a direction normal to the phase boundary (i.e. in
the notation of Section \ref{motiv}, normal to $\cal{P}$). If the
band now moves with a velocity $\bfV^S$ relative to the material,
at a material point past which the boundary moves there is change
of i-elastic distortion per unit time given by $\bfSigma$. The
geometrically complete local evolution equation for $\bfS$ is
given by
\begin{equation}\label{S_evol}
{\stackrel{\diamond}{\bfS}} = - \bfPi \times \bfV^{\Pi} + grad \, \left( \bfW \bfS^{2T}\bfV^S \right).
\end{equation}

\begin{figure}
\begin{center}
\includegraphics[height=0.1\textheight,width=0.9\textwidth,angle=0]{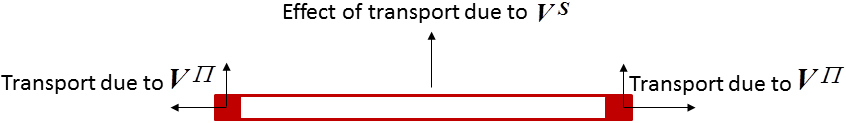}
\caption{Transport due to g.disclination and eigenwall velocities.}
\label{velocities}
\end{center}
\end{figure}

Thus, for phase boundaries, $\bfV^{\Pi}$ transports in-plane
gradients of $\bfS$ including the tips of such bands, whereas
$\bfV^S$ transports the phase boundary transverse to itself
(Figure \ref{velocities}).
\subsection{Dislocation density and i-elastic 1-distortion evolution}\label{disloc_density_evol}
Following tradition \cite{dewit_disclinations_II}, we define the \emph{dislocation density}  $\bfalpha$ as
\begin{equation}\label{alpha_def}
\bfalpha := \bfY:\bfX = \left(\bfS + grad\, \bfW \right):\bfX
\end{equation}
and note that when $\bfS \equiv \bf0$, $\bfalpha = - curl\,\bfW$
since for any smooth tensor field $\bfA$, $curl\, \bfA =
-grad\,\bfA :\bfX$. The definition (\ref{alpha_def}) is motivated
by the displacement jump formula (\ref{disp_jump}) corresponding
to a single, isolated defect line terminating an i-elastic
distortion jump in the body. In this situation, the displacement
jump for an isolated defect line, measured by integrating
$\bfalpha$ on an area patch threaded by the defect line, is no
longer a topological object independent of the area patch.

The evolution of the $\bfS:\bfX$ component of $\bfalpha$ is
already specified from the evolution (\ref{S_evol}) for $\bfS$.
Thus, what remains to be specified for the evolution of the
dislocation density field is the evolution of
\[
\widetilde{\bfalpha} := -curl\,\bfW = \left(\bfY - \bfS
\right):\bfX,
\]
that is again an areal density of lines carrying a vectorial attribute.

\emph{When} $\bfS = \bf0$, then $\widetilde{\bfalpha} = \bfalpha$,
and the physical arguments of finite-deformation dislocation
mechanics \cite{acharya2011microcanonical} yield
\[
\dot{ \overline{ \int_{a(t)} \widetilde{\bfalpha}\,  \bfn \, da }}
  = - \int_{c(t)} \widetilde{\bfalpha} \times \bfV^{\alpha} \, d\bfx
\]
with corresponding local form
\[
\dot{\bfW} + \bfW\bfL = \widetilde{\bfalpha} \times \bfV^{\alpha},
\]
(up to assuming an additive gradient of a vector field to vanish).
Here, $\bfV^{\alpha}$ denotes the \emph{dislocation velocity}
field, to be interpreted physically as the field responsible for
transporting the dislocation density field in the body.

Using identical logic, we assume as the statement of evolution of $\bfW$ the equation
\begin{equation}\label{W_evolve}
\dot{\bfW} + \bfW\bfL = \bfalpha \times \bfV^{\alpha},
\end{equation}
with a natural adjustment to reflect the change in the definition
of the dislocation density field. This statement also corresponds
to the following local statement for the evolution of
$\widetilde{\bfalpha}$:
\begin{equation}\label{alphahat_evolve}
\stackrel{\circ}{\widetilde{\bfalpha}} := \left( div \, \bfv \right) \widetilde{\bfalpha} + \dot{\widetilde{\bfalpha}} - \widetilde{\bfalpha} \bfL^T = -curl \left( \bfalpha \times \bfV^\alpha \right).
\end{equation}

It is to be noted that in this generalization of the
dislocation-only case, the dislocation density is no longer
necessarily divergence-free (see (\ref{alpha_def})) which is
physically interpreted as the fact that dislocation lines may
terminate at eigenwalls or phase boundaries.

We note here that (\ref{W_evolve}) can be rewritten in the form
\begin{equation}\label{add_decomp}
\bfL = \dot{\bfF}^e\bfF^{e-1} + \left( \bfF^e \bfalpha \right) \times \bfV^{\alpha},
\end{equation}
where $\bfF^e := \bfW^{-1}$. To make contact with classical finite
deformation elastoplasticity, this may be interpreted as a
fundamental additive decomposition of the velocity gradient into
elastic $\left( \dot{\bfF}^e\bfF^{e-1} \right)$ and plastic
$\left( \left( \bfF^e \bfalpha \right) \times \bfV^{\alpha}
\right)$ parts. The latter is defined by the rate of deformation
produced by the flow of dislocation lines in the current
configuration, \emph{without any reference to the notion of a
total plastic deformation from some pre-assigned reference
configuration}. We also note the \emph{natural emergence of
plastic spin} (i.e. a non-symmetric plastic part of $\bfL$), even
in the absence of any assumptions of crystal structure but arising
purely from the kinematics of dislocation motion (when a
dislocation is interpreted as an elastic incompatibility). 
\subsection{Summary of proposed mechanical structure of the theory}
To summarize, the governing equations of the proposed model are
\begin{eqnarray}\label{gov_eqns}\nonumber
\dot{\rho}  & = & - \rho \divergence \bfv\\\nonumber
\rho\dot{\bfv} &= & \divergence \bfT + \rho\bfb  \\
\bf0  & = & \divergence \bfLambda - \bfX:\bfT + \rho\bfK \\\nonumber
\dot{\bfW}  & = & - \bfW\bfL + \bfalpha \times \bfV^{\alpha}\\\nonumber
\dot{\bfS} &  = & \bfW^{-1} \left\{ - \dot{\bfW} \bfS^{2T} - \bfW\bfS^{2T}\bfL - \bfPi \times \bfV^{\Pi} + grad \, \left( \bfW \bfS^{2T}\bfV^S \right) \right\}^{2T} \\\nonumber
\bf0 & = & - \bfalpha + \bfS:\bfX - curl\, \bfW.
\end{eqnarray}
The fundamental dependent fields governed by these equations are
the current position field $\bfx$, the i-elastic 1-distortion
field $\bfW$, and the eigenwall field $\bfS$.

The relevance of the eigenwall velocity field $\bfV^S$ would seem
to be greatest in the completely compatible case when there are no
deformation line defects allowed (i.e. $\bfalpha = \bf0$, $\bfPi =
\bf0$). For reasons mentioned in section (\ref{compatible}),
including eigenwall evolution seems to be at odds with strict
compatibility. Additionally, modeling wall defects by dipolar
arrays of disclinations \cite{taupin_shear_coupled} appears to be
a successful, fundamental way of dealing with grain boundary
motion. However, it also seems natural to consider many phase
boundaries as containing no g.disclinations whatsoever, e.g. the
representation of a straight phase boundary of constant strength
that runs across the body without a termination (this may be
physically interpreted as a consistent coarser length-scale view
of a phase-boundary described by separated g.disclination-dipole
units). To represent phase boundary motion in this situation of no
disclinations, a construct like $\bfV^S$ is necessary, and we
therefore include it for mathematical completeness.

The model requires constitutive specification for
\begin{itemize}
\item the stress $\bfT$,
\item the couple-stress $\bfLambda$,
\item the g.disclination velocity $\bfV^{\Pi}$,
\item the dislocation velocity $\bfV^{\alpha}$, and
\item the eigenwall velocity $\bfV^S$ (when not constrained to vanish).
\end{itemize}

As a rough check on the validitiy of the mechanical structure, we
would like to accommodate analogs of the following limiting model
scenarios within our general theory. The first corresponds to the
calculation of static stresses of disclinations in linear
elasticity \cite{dewit_disclinations_II}, assuming no dislocations
are present. That is, one thinks of a terminating surface of
discontinuity in the elastic rotation field, across which the
elastic displacements are continuous (except at the singular tip
of the terminating surface). The analog of this question in our
setting would be to set $\bfalpha = \bf0$ in (\ref{alpha_def}) and
consider $\bfS : \bfX$ as a given source for $\bfW$, i.e.
\[
\widetilde{\bfalpha} = - curl \, \bfW = - \bfS : \bfX,
\]
where $\bfW$ is assumed to be the only argument of the stress
tensor. Thus, the $\bfS$ field directly affects the elastic
distortion that feeds into the stress tensor. Of course, this
constrained situation, i.e. $\bfalpha = \bf0$, may only be
realized if the field $\bfS:\bfX$ is divergence-free on $\cal{B}$.
Thus, with (\ref{alpha_def}) as a field equation along with
constitutive equations for the stress and couple stress tensor and
the static versions of balance of linear and angular momentum,
this problem becomes accessible within our model.

As a second validating feature of the presented model, we mention
the work of \cite{taupin_shear_coupled} on the prediction of shear
coupled grain boundary migration within what may be interpreted as
a small-deformation, disclination-dislocation-only version of the
above theory. There, the grain boundaries are modeled by an array
of (stress-inducing) disclination dipoles and it is shown how the
kinematic structure of the above type of system along with the
presence of stresses and couple stresses allows grain boundary
motion with concomitant shear-producing dislocation glide to be
predicted in accord with experiments and atomistic simulations.

Finally, one would of course like to recover some regularized
version of classical, compatible phase transformation theory
\cite{ball1989fine}, i.e. classical nonlinear elasticity with a
non-convex energy function and with continuous displacements, in
the absence of dislocations, g.disclinations and the eigenwall
field in our model, i.e. $\left( \bfalpha = \bf0, \bfS = \bfPi =
\bf0 \right)$. The model reduces to a strain gradient
regularization \cite{slemrod1983admissibility,
abeyaratne1991implications, barsch1984twin, shenoy1999martensitic}
of classical nonlinear elasticity resulting from the presence of
couple stresses and the dependence of the energy function on the
second deformation gradient. 
\subsection{The possibility of additional kinetics in the completely compatible case}\label{compatible}
The question of admitting additional kinetics of phase boundary
motion in the completely compatible case (i.e. no dislocations and
g.disclinations) is an interesting one, raised in the works of
Abeyaratne and Knowles \cite{abeyaratne1990driving,
abeyaratne1991implications}. In the spatially 1-d scenario
considered in \cite{abeyaratne1991implications}, it is shown that
admitting higher gradient effects does provide additional
conditions over classical elasticity for well-defined propagation
of phase boundaries, albeit with no dissipation, while the results of \cite{slemrod1983admissibility} show that a viscosity effect alone is too restrictive and does not allow propagation. The work of \cite{abeyaratne1991implications}, that extends to 3-d \cite{abeyaratne2006evolution}, does
not rule out, and in fact emphasizes, more general kinetic
relations for phase boundary propagation arising from dissipative effects, demonstrating the fact through a combined viscosity-capillarity regularization of nonlinear elasticity.

Within our model, the analogous situation is to consider the
g.disclination density and the dislocation density constrained to
vanish ($\bfPi = \bf0$ and $\bfalpha =  \bf0$). A dissipative
mechanism related to phase boundary motion may now be introduced
by admitting a generally non-vanishing $\bfV^S$ field. For the
present purpose, it suffices then to focus on the following three
kinematic equations:
\[
\dot{\bfS}  = \bfW^{-1} \left\{ - \dot{\bfW} \bfS^{2T} - \bfW\bfS^{2T}\bfL + grad \, \left( \bfW \bfS^{2T}\bfV^S \right) \right\}^{2T}
\]
\begin{equation}\label{comp_kin}
\dot{\bfW}   =  - \bfW\bfL + \bfalpha \times \bfV^{\alpha}
\end{equation}
\[
\bf0  =  - \bfalpha + \bfS:\bfX - \mit{curl} \, \bfW.
\]
We first note from (\ref{comp_kin}$_2$) that if $\bfalpha = \bf0$
then a solution for $\bfW$ with initial condition $\bfI$ would be
$\bfF^{-1}$, where $\bfF$ is the deformation gradient with respect
to the fixed stress-free reference configuration. Then from
(\ref{comp_kin}$_3$), it can be seen that this ansatz requires the
eigenwall field to be symmetric in the last two indices. In its
full-blown geometric nonlinearity, it is difficult to infer from
(\ref{comp_kin}$_1$) that if $\bfS$ were to have initial
conditions with the required symmetry, that such symmetry would
persist on evolution.

An even more serious constraint within our setting making
additional kinetics in the completely compatible case dubious is
the further implication that if $\bfPi = curl\, \bfS = \bf0$ and
$\bfS:\bfX = \bf0$ on a simply connected domain, then it is
necessarily true that $\bfS$ can be expressed as the second
gradient of a vector field say $\bfa$, i.e.
\begin{equation}\label{S_comp}
S_{ijk} = a_{i,jk}.
\end{equation}
This implies that (\ref{comp_kin}$_1$) is in general a highly
overdetermined system of $27$ equations in $3$ unknown fields, for
which solutions can exist, if at all, for very special choices of
the eigenwall velocity field $\bfV^S$. Even in the simplest of
circumstances, consider (\ref{comp_kin}$_1$) under the
geometrically linear assumption (i.e. all nonlinearities arising
from an objective rate are ignored and we do not distinguish
between a material and a spatial time derivative)
\[
\dot{\bfS}  =  grad \, \left( \bfS \bfV^S \right) \Longrightarrow \dot{a}_{i,j} = a_{i,jk} \left( V^S \right)_k
\]
(upto a spatially uniform tensor field). This is a generally
over-constrained system of $9$ equations for $3$ fields
corresponding to the evolution of the vector field $\bfa$
requiring, for the existence of solutions, a pde constraint to be
satisfied by the phase boundary/eigenwall velocity field, namely
\[
curl \, \left\{ \left( grad \, grad \, \bfa \right) \bfV^S \right\} = \bf0
\]
that amounts to requiring that
\[
a_{i,jk} \left( V^S_{k,l} \right) - a_{i,lk} \left( V^S_{k,j} \right) = 0.
\]
While satisfied in some simple situations, e.g. $grad \, \bfV^S =
\bf0$ wherever $grad \, grad \, \bfa$ is non-vanishing, or when
all field-variations are in one fixed direction (as for phase
boundary propagation in a 1-d bar), this is a non-trivial
constraint on the $\bfV^S$ field in general. Of course, it is
conventional wisdom that the phase boundary velocity kinetics be
specifiable constitutively, and a `nonlocal' constraint on
$\bfV^S$ as above considerably complicates matters. On the other
hand, we find it curious that a nonlocal constraint on phase
transformation constitutive behavior arises naturally in our model
as a consequence of enforcing strict kinematic compatibility.

If one disallows a non-local PDE constraint as above on the
constitutive specification of $\bfV^S$, then the kinematics
suggests the choice $\bfV^S = \bf0$ (and perhaps the even stronger
$\dot{\bfS} = \bf0$). Based on the results of Section
\ref{sec_rev_diss}, this \emph{precludes dissipation in the
completely compatible case}. We find it interesting that recent
physical results guided by continuum mechanics theory
\cite{cui2006combinatorial, zarnetta2010identification} point to a
similar conclusion in the design of low-hysteresis
phase-transforming solids. 
\subsection{Contact with the classical view of modeling defects: A Weingarten theorem for g.disclinations and associated dislocations}
\begin{figure}
\begin{center}
\includegraphics[height=0.4\textheight,width=0.8\textwidth,angle=0]{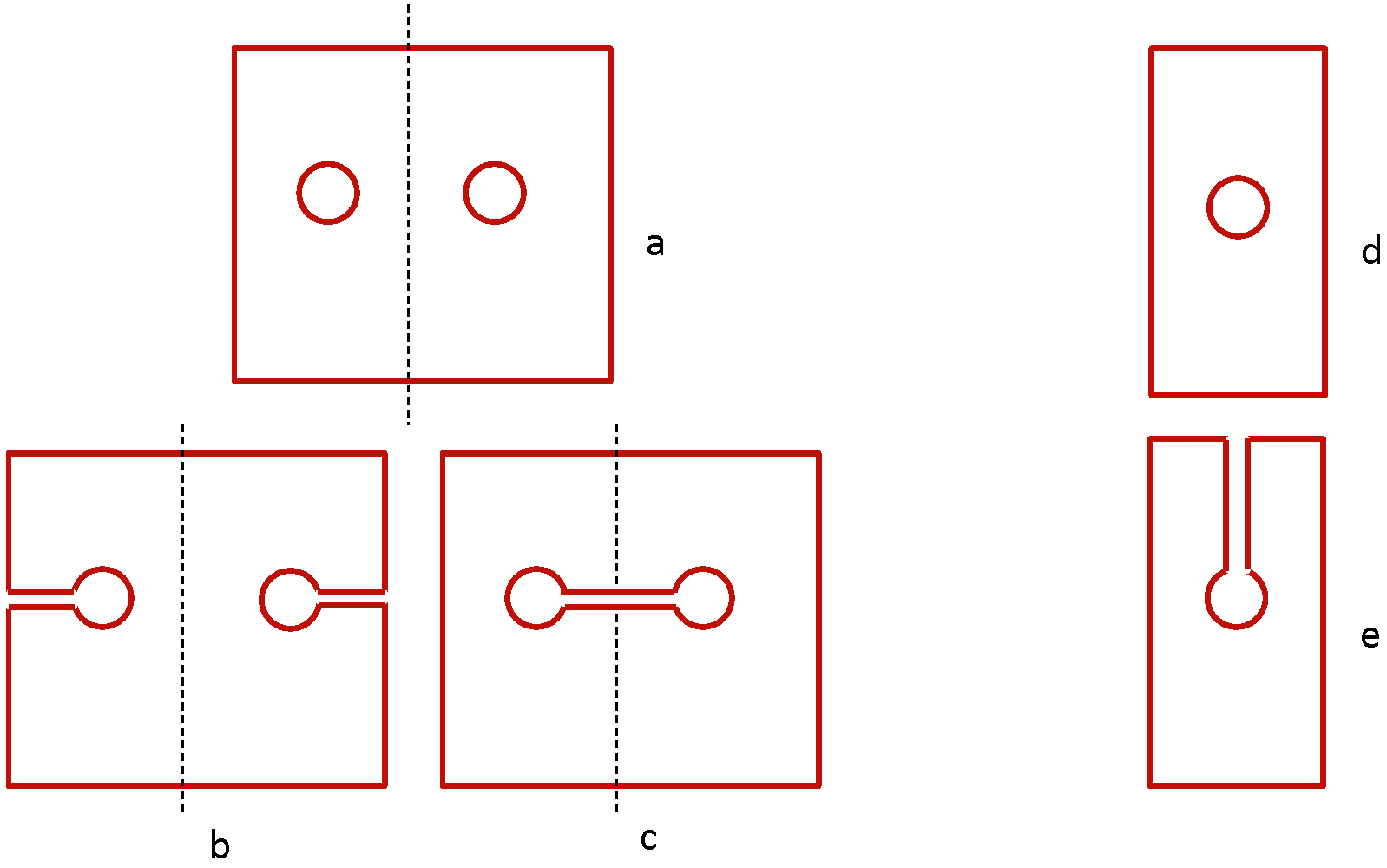}
\caption{Non simply-connected and corresponding induced
simply-connected bodies. For a,b,c the bodies are obtained by
rotating the planar figures by $\pi$ about the axes shown; for
d,e they are obtained by extruding the planar figures along the axis perpendicular to the plane of the paper.}
\label{multiply_connected}
\end{center}
\end{figure}

The discussion surrounding (\ref{S_comp}) and seeking a connection
of our work to the classical tradition of the theory of isolated
defects suggest the following natural question. Suppose one has a
three-dimensional body with a toroidal (Figure
\ref{multiply_connected}a) or a through-hole in it (Figure
\ref{multiply_connected}d)(cf. \cite{nabarro1967theory}). In both
cases, the body is multiply-connected. In the first, the body can
be cut by a surface of finite extent that intersects its exterior
surface along a closed curve and the surface of the toroidal hole
along another closed curve in such a way that the resulting body
becomes simply-connected with the topology of a solid sphere
(Figure \ref{multiply_connected}b). In more precise terminology,
one thinks of isolating a surface of the original
multiply-connected domain with the above properties, and the set
difference of the original body and the set of points constituting
the cut-surface is the resulting simply-connected domain induced
by the cut. Similarly, the body with the through-hole can be cut
by a surface extending from a curve on the external surface to a
curve on the surface of the through-hole such that the resulting
body is again simply-connected with the topology of a solid sphere
(Figure \ref{multiply_connected}e). Finally, the body with the
toroidal hole can also be cut by a surface bounded by a closed
curve entirely on the surface of the toroidal hole in such a way
that the resulting body is simply-connected with the topology of a
solid sphere with a cavity in it. For illustration see (Figure
\ref{multiply_connected}c).

To make contact with our development in Section \ref{motiv}, one
conceptually associates the support of the defect core as the
interior of the toroidal hole and the support of the strip field
$\bfW\bfV$ as a regularized cut-surface.

Suppose that on the original multiply-connected domain
\begin{itemize}
\item a continuously differentiable, $3^{rd}$-order tensor field $\tilde{\bfY}$ is prescribed that is
\item symmetric in its last two indices $( \tilde{Y}_{ijk} = \tilde{Y}_{ikj} )$ and
\item whose $curl$ vanishes $( \tilde{Y}_{ijk,l} = \tilde{Y}_{ijl,k})$.
\end{itemize}
Given such a field, we ask the question of whether on the
corresponding simply-connected domain induced by a cut-surface as
described in the previous paragraph, a vector field $\bfy$ can be
defined such that
\[
grad\,grad \, \bfy = \tilde{\bfY} \ \ \ ; \ \ \ y_{i,jk} = \tilde{Y}_{ijk},
\]
and if the difference field of the limiting values of $\bfy$, as
the cut-surface is approached from the two sides of the body
separated by the cut-surface, i.e. the jump $\llbracket \bfy
\rrbracket$ of $\bfy$ across the cut, is arbitrary or yields to
any special characterization. Here, we will refer to limits of
fields approached from one (arbitrarily chosen) side of the
cut-surface with a superscript `$+$' and limits from the
corresponding other side of the cut-surface with a superscript
`$-$' so that, for instance, $\llbracket \bfy(\bfz) \rrbracket =
\bfy^+(\bfz) - \bfy^-(\bfz)$, for $\bfz$ belonging to the
cut-surface.

For the question of existence of $\bfy$ on the simply-connected domain, one first looks for a field $\tilde{\bfW}$ such that
\[
grad\, \tilde{\bfW} = \tilde{\bfY} \,\ \ \ ; \ \ \  \tilde{W}_{ij,k} = \tilde{Y}_{ijk}
\]
and since $\tilde{\bfY}$ is $curl$-free and continuously
differentiable on the multiply-connected domain with the hole, on
the corresponding simply-connected domain induced by a cut, the
field $\tilde{\bfW}$ can certainly be defined
\cite{thomas1934systems}. The jump $\llbracket \tilde{\bfW}
\rrbracket$ is not to be expected to vanish on the cut surface, in
general. However, by integrating $\left(
grad\,\tilde{\bfW}\right)^+$ and $\left(
grad\,\tilde{\bfW}\right)^-$ along a curve on the cut-surface
joining any two arbitrarily chosen points on it, it is easy to
deduce that $\llbracket \tilde{\bfW} \rrbracket$ is constant on
the surface because of the continuity of $\tilde{\bfY}$ on the
original multiply-connected domain.

\begin{figure}
\begin{center}
\includegraphics[height=0.4\textheight,width=0.6\textwidth,angle=0]{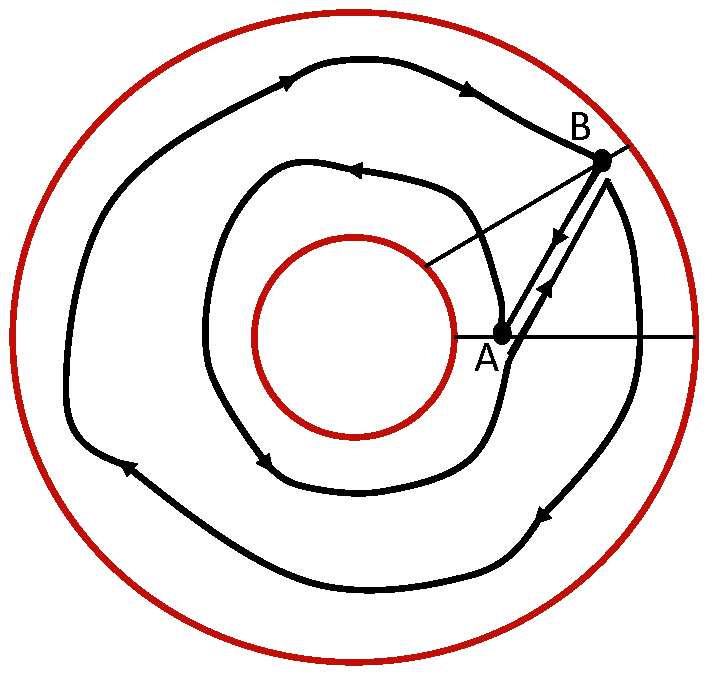}
\caption{Contour for proving independence of $\bfDelta$ on
cut-surface. The contour need not be planar and the points $A$ and
$B$ need not be on the same cross-sectional plane of the body.}
\label{Delta}
\end{center}
\end{figure}

With reference to (Figure \ref{Delta}),  consider the line
integral of $\tilde{\bfY}$ on the closed contour shown in the
original multiply-connected domain without any cuts (the two
oppositely-oriented adjoining parts of the contour between points
$\bfA$ and $\bfB$ are intended to be overlapping). In conjunction,
also consider as the `inner' and `outer' closed contours the
closed curves that remain by ignoring the overlapping segments,
the inner closed contour passing through $\bfA$ and the outer
through $\bfB$. Then, because of the continuity of $\tilde{\bfY}$
and its vanishing $curl$,  the line integral of $\tilde{\bfY}$ on
the inner and outer closed contours must be equal and this must be
true for \emph{any} closed circuit that cannot be shrunk to a
point while staying within the domain. Let us denote this
invariant over any such closed curve $\cal{C}$ as
\[
\int_{\cal{C}} \tilde{\bfY}\,d\bfx = \bfDelta.
\]
If we now introduce a cut-surface passing through $\bfA$ and
construct the corresponding $\tilde{\bfW}$, say $\tilde{\bfW}_1$,
then the jump of $\tilde{\bfW}_1$ at $\bfA$ is given by
\[
\llbracket \tilde{\bfW}_1 \rrbracket (\bfA) = \int_{\cal{C}({\bfA^-},{\bfA^+})} grad\, \tilde{\bfW}_1 \, d \bfx = \int_{\cal{C}({\bfA^-},{\bfA^+})} \tilde{\bfY} \, d \bfx  = \bfDelta,
\]
where $\cal{C}(\bfA^-,\bfA^+)$ is the curve formed from the inner
closed contour defined previously with the point $\bfA$ taken out
and with start-point $\bfA^-$ and end-point $\bfA^+$. The last
equality above is due to the continuity of $\tilde{\bfY}$ on the
original multiply-connected domain. Similarly, a different
cut-surface passing through $\bfB$ can be introduced and an
associated $\tilde{\bfW}_2$ constructed with $\llbracket
\tilde{\bfW}_2 \rrbracket (\bfB) = \bfDelta$. Since $\bfA$, $\bfB$
and the cut surfaces through them were chosen arbitrarily, it
follows that the jump of any of the functions $\llbracket
\tilde{\bfW} \rrbracket$ across their corresponding cut-surface
takes on the same value regardless of the cut-surface invoked to
render simply-connected the multiply-connected body.

On a cut-induced simply-connected domain, since $\tilde{\bfW}$
exists and its $curl$ vanishes (due to the symmetry of
$\tilde{\bfY}$ in its last two indices), clearly a vector field
$\bfy$ can be defined such that
\[
grad \,\bfy = \tilde{\bfW}.
\]
Suppose we now fix a cut-surface. Let $\bfx_0$ be an arbitrarily
chosen base point on it. Let $\bfx$ be any other point on the
cut-surface. Then, by integrating $\left(grad\,\bfy \right)^+$ and
$\left(grad\,\bfy \right)^-$ along any curve lying on the
cut-surface joining $\bfx_0$ and $\bfx$, it can be observed that
\begin{equation}\label{disp_jump}
\llbracket \bfy(\bfx) \rrbracket = \llbracket \bfy(\bfx_0) \rrbracket + \bfDelta \left( \bfx - \bfx_0 \right).
\end{equation}
The `constant vector of translation', $\llbracket \bfy(\bfx_0)
\rrbracket$, may be evaluated by integrating $\tilde{\bfW}$ on a
closed contour that intersects the cut-surface only once, the
point of intersection being the base point $\bfx_0$
($\tilde{\bfW}$ is, in general, discontinuous at the base point).
It can be verified that for a fixed cut-surface, $\llbracket
\bfy(\bfx) \rrbracket$ is independent of the choice of the base
point used to define it.

The physical result implied by this characterization is as
follows: suppose we think of the vector field $\bfy$ as a
generally discontinuous deformation of the multiply-connected
body, with discontinuity supported on the cut-surface. Then the
separation/jump vector $\bfy(\bfx)$ for \emph{any} point $\bfx$ of
the surface corresponds to a \emph{fixed}  affine deformation of
the position vector of $\bfx$ relative to the base point $\bfx_0$
(i.e. $\bfDelta\,$ independent of $\bfx$), followed by a
\emph{fixed} translation.

It is important to note here that, for the given field
$\tilde{\bfY}$ on the multiply-connected domain, while $\bfDelta =
\llbracket \tilde{\bfW} \rrbracket$ is independent of the
particular cut-surface invoked to define it, the translational
part, $\llbracket \bfy(\bfx_0) \rrbracket$, of the jump
$\llbracket \bfy \rrbracket$  on a cut-surface depends on the
definition of the cut-surface (both through the dependence on
$\bfx_0$ and the impossibility, in general, of defining a
continuous $\tilde{\bfW}$ on the original multiply-connected
domain), \emph{unless} $\bfDelta = \bf0$. This is the analog of
the known result in classical (disclination-dislocation) defect
theory that the Burgers vector of an isolated defect is a
well-defined topological object only in the absence of
disclinations. In the same spirit, when the (non-trivial) constant
tensor $\bfDelta$ is such that it has a 2-dimensional null-space,
then for a specific flat, cut-surface spanning the null-space, it
is possible that the jump in $\llbracket \bfy \rrbracket$
vanishes. This gives rise to a surface in the
(non-simply-connected) body on which the deformation map is
continuous but across which the deformation gradient is
discontinuous.

Thus, the notion of g.disclinations offers more flexibility in the
type of discontinuities that can be represented within continuum
theory, as compared to Volterra distortions defining classical
disclinations (cf. \cite{casey2004volterra,
nabarro1967theory})\footnote 
{In the classical disclination-dislocation case, the corresponding
question to what we have considered would be to ask for the
existence, on a cut-induced simply-connected domain, of a vector
field $\bfy$ and the characterization of its jump field across the
cut-surface, subject to $(grad\,\bfy)^T\grad\,\bfy = \bfC$ and the
Riemann-Christoffel curvature tensor field of  (twice continuously
differentiable) $\bfC$ (see \cite{shield1973rotation} for
definition) vanishing on the original multiply-connected domain.
Existence of a global smooth solution can be shown (cf.
\cite{sokolnikoff1951tensor} using the result of
\cite{thomas1934systems} and the property of preservation of
inner-product of two vector fields under parallel transport in
Riemannian geometry). The corresponding result is
\[
\llbracket \bfy(\bfx) \rrbracket = \llbracket \bfy(\bfx_0) \rrbracket + \llbracket \bfR \rrbracket \bfU \left( \bfx - \bfx_0 \right),
\]
where $grad\,\bfy = \bfR \bfU$ on the cut-induced simply-connected
domain, and $\bfR$ is a proper-orthogonal, and $\bfU =
\sqrt{\bfC}$ is a symmetric, positive-definite, $2^{nd}$-order
tensor field. $\bfU$ cannot have a jump across any cut-surface and
the jump $\llbracket \bfR \rrbracket$ takes the same value
regardless of the cut-surface invoked to define it, as can be
inferred from the results of \cite{shield1973rotation}. By
rearranging the independent-of-$\bfx$ term in the above
expression, the result can be shown to be identical to that in
\cite{casey2004volterra}. Of course, for the purpose of
understanding the properties of the Burgers vector of a general
defect curve, it is important to observe the dependence of the
`constant' translational term on the cut-surface. An explicit
characterization of the jump in $grad\, \bfy$ in terms of the
strength of the disclination is given in
\cite{derezin2011disclinations}.}. 
This is natural since the Volterra distortion question involves a
twice-continuously differentiable Right-Cauchy Green field in its
formulation (in the context of this subsection, this would amount
to enforcing a high degree of smoothness, and therefore
continuity, on $\tilde{\bfW}^T\tilde{\bfW}$) so that only the
polar decomposition-related rotation field of $\tilde{\bfW}$ can
be discontinuous, whereas allowing for an incompatible
$\tilde{\bfY}$ field on a multiply-connected domain, even though
irrotational, implies possible discontinuities in the whole field
$\tilde{\bfW}$. 

\section{Frame-Indifference and thermodynamic guidance on constitutive structure}\label{MFI_thermo}

As is known to workers in continuum mechanics, the definition of
the mechanical dissipation (\ref{eqn:dissipation}) coupled to the
mechanical structure of a theory (Section \ref{mech_structure}), a
commitment to constitutive dependencies of the specific
free-energy density, and the consequences of material frame
indifference provide an invaluable tool for discovering the
correct form of the reversible response functions and driving
forces for dissipative mechanisms in a nonlinear theory. This
exercise is useful in that constitutive behavior posed in
agreement with these restrictions endow the theory with an energy
equality that is essential for further progress in developing
analytical results regarding well-posedness as well as developing
numerical approximations. In exploiting this idea for our model,
we first deduce a necessary condition for frame-indifference of
the free-energy density function that we refer to as the `Ericksen
identity' for our theory; in this, we essentially follow the
treatment of \cite{anderson1999continuum} adapted to our context.

\subsection{Ericksen identity for g.disclination mechanics}

We assume a specific free energy density of the form
\begin{equation}\label{free_energy}
\psi = \psi\left( \bfW, \bfS, \bfJ, \pbPibf \right).
\end{equation}
All the dependencies above are two-point tensors between the
current configuration and the `intermediate configuration,' i.e.
$\left\{ \calD_r(\bfx,t): \bfx \in B(t) \right\}$, a collection of
local configurations with similarly oriented and unstretched
atomic configurations in each of them. On superimposing rigid
motions on a given motion, each element of this intermediate
configuration is naturally assumed to remain invariant. With this
understanding, let $\bfQ(s)$ is a proper-orthogonal tensor-valued
function of a real parameter $p$ defined by
\[
\frac{d \bfQ}{dp} (p) = \bfs \bfQ (p),
\]
where $\bfs$ is an arbitrarily fixed skew-symmetric tensor
function, and $\bfQ(0) = \bfI$. Thus, $\frac{d \bfQ^T}{dp} (0) = -
\bfs$. Also, define $\bfA \, t \bfB$ through
\[
\left\{ \left(A\, t B)_{jkrl} - A_{jr}B_{kl} \right) \right\} \bfe_j \otimes \bfe_k \otimes \bfe_r \otimes \bfe_l = \bf0.
\]
Then, frame-indifference of $\psi$ requires that
\begin{equation}\label{psi_indiff}
\hpsi \left( \bfW, \bfS, \bfJ, \pbPibf \right) = \hpsi \left( \bfW \bfQ^T, \bfS :\bfQ^T t \bfQ^T, \bfJ :\bfQ^T t \bfQ^T, \pbPibf \bfQ^T \right)
\end{equation}
for $\bfQ(p)$ generated from any choice of the skew symmetric
tensor $\bfs$. Differentiating (\ref{psi_indiff}) with respect to
$p$ and evaluating at $p = 0$ implies
\begin{eqnarray}\nonumber
0  & = & -\left( \partial_{\bfW} \hpsi \right)_{ij} W_{ir} s_{rj} - \left( \partial_{\bfS} \hpsi \right)_{ijk} S_{irs} \left( s_{rj} \delta_{sk} + \delta_{rj} s_{sk} \right) \\\nonumber
 &  & - \left( \partial_{\bfJ} \hpsi \right)_{ijk} J_{irs} \left( s_{rj} \delta_{sk} + \delta_{rj} s_{sk} \right) - \left( \partial_{\pbPibf} \hpsi \right)_{ijk} \pbPi_{ijr} s_{rk}
\end{eqnarray}
where the various partial derivatives of $\hpsi$ are evaluated at $\left( \bfW, \bfS, \bfJ, \pbPibf \right)$. This can be rewritten as
\begin{eqnarray}\label{ericksen}
0 & = & \left[ \left( \partial_{\bfW} \hpsi \right)_{ij} W_{ir}  \right.  \ \ + \left( \partial_{\bfS} \hpsi \right)_{ijk} S_{irk} + \left( \partial_{\bfS} \hpsi \right)_{ikj} S_{ikr}\\\nonumber
 &  & \ \ + \left( \partial_{\bfJ} \hpsi \right)_{ijk} J_{irk} + \left( \partial_{\bfJ} \hpsi \right)_{ikj} J_{ikr} + \left. \left( \partial_{\pbPibf} \hpsi \right)_{ikj} \pbPi_{ikr} \right]\,s_{rj},
\end{eqnarray}
valid for all skew symmetric $\bfs$ which implies that the term
within square brackets has to be a symmetric second-order tensor.
\emph{This is a constraint on constitutive structure imposed by
Material Frame Indifference}.

\subsection{The mechanical dissipation}\label{mech_diss}
Assuming a stored energy density function $\hpsi$ with arguments
as in (\ref{free_energy}), we now re-examine the mechanical
dissipation $\textsf{D}$ in (\ref{eqn:dissipation}). We first
compute the material time derivative of $\hpsi$ to obtain
\begin{equation}\label{psidot}
\begin{split}
\dot{\psi} = \delpsi{\bfW}: & \dot{\bfW} + \delpsi{\bfS} \cdot_3 \, \dot{\bfS} + \delpsi{\bfJ} \cdot_3 \, \dot{\bfJ} + \delpsi{\pbPibf} \cdot_3 \, \dot{\pbPibf} \\
= \delpsi{\bfW}: & \left( \right. -\bfW \bfL + \bfalpha \times \bfV^\alpha \left. \right) \\
 + \delpsi{\bfS} & \cdot_3 \left( \right.  \bfW^{-1} \left\{ \right. - \dot{\bfW} \bfS^{2T} - \bfW\bfS^{2T}\bfL - \bfPi \times \bfV^{\Pi}
 \\
& \ \ \ \ \ \ + \, grad \, \left( \right.  \bfW \bfS^{2T}\bfV^S \left. \right) \left. \right\}^{2T} \left. \right)\\
 + \delpsi{\bfJ} & \cdot_3 \dot{\bfJ} \\
 + \delpsi{\pbPibf} & \cdot_3 \left[ \right. - (\bfL : \bfI) \pbPibf + \pbPibf {\bfL}^T - curl \, \left( \bfPi \times \bfV^\Pi \right) \left. \right].
\end{split}
\end{equation}
In the above, $\cdot_3$ refers to the inner-product of its
argument third-order tensors (in indices, a contraction on all
three (rectangular Cartesian) indices of its argument tensors).
Recalling the dissipation (\ref{eqn:dissipation}):
\[
\textsf{D} = \int_{B(t)} \left( \bfT:\bfD + \bfLambda: \bfM - \rho\dot{\psi} \right) \ dv,
\]
we first collect all terms in (\ref{psidot}) multiplying $\bfL =
\bfD + \bfOmega$ and $grad \, \bfL$, observing that the
coefficient of $\bfOmega$ has to vanish identically for the
dissipation to be objective (cf. \cite{acharya-dayal}). Noting
that
\[
\dot{\bfJ} = grad \, \dot{\bfW} - \left( grad \, \bfW \right) \bfL \Longleftrightarrow \dot{\overline{W_{rw,k}}} = \left(\dot{W_{rw}}\right)_{,k} - W_{rw,m}L_{mk},
\]
we obtain
\begin{equation*}
\begin{split}
\int_{B(t)} \left[ \right. & - \delpsi{\bfW}_{ij}W_{ir} L_{rj} \\
& + \delpsi{\bfS}_{rwk} W^{-1}_{wl} \left( W_{lr}L_{rp}S_{rpk} - W_{lp}S_{rpm}L_{mk} \right)\\
& + \delpsi{\bfJ}_{rwk} \left( -W_{rp,k}L_{pw} - W_{rw,m}L_{mk} \right)\\
&  + \delpsi{\pbPibf}_{ikj} \left( \pbPi_{ikr}L^T_{rj} - L_{rr}\pbPi_{ikj} \right) \left. \right] \, dv \\
& + \, \int_{B(t)} \delpsi{\bfJ}_{rwk} \left( - W_{rp} L_{pw,k} \right) \, dv,
\end{split}
\end{equation*}
Noting the symmetry of $L_{pwk}$ in the last two indices we \emph{define}
\[
\left( D_{\bfJ}^{sym} \psi \right)_{rwk} := \frac{1}{2} \left[ \delpsi{\bfJ}_{rwk} +  \delpsi{\bfJ}_{rkw} \right],
\]
and substituting the above in the dissipation
(\ref{eqn:dissipation}) to collect terms `linear' in $\bfD$,
$\bfOmega$, and $grad\,\bfOmega$, we obtain
\begin{equation}\label{reversible}
\begin{split}
- \int_{B(t)} - \rho \left[ \right. & \delpsi{\bfW}_{ij}W_{ir} + \delpsi{\bfS}_{ijk} S_{irk} + \delpsi{\bfS}_{ikj} S_{ikr} \\
 & + \delpsi{\bfJ}_{ijk} J_{irk} + \delpsi{\bfJ}_{iwj} J_{iwr} \\
 & + \delpsi{\pbPibf}_{ikj}\pbPi_{ikr} \left. \right]\, \Omega_{rj} \, dv\\
+ \int_{B(t)} \left[ \right. T_{rj} - \rho \left\{ \right.  & - \delpsi{\bfW}_{ij}W_{ir} \\
& + \delpsi{\bfS}_{mrk} S_{mjk} - \delpsi{\bfS}_{mwj} S_{mwr} \\
& - \delpsi{\bfJ}_{pjk} J_{prk} - \delpsi{\bfJ}_{mwj} J_{mwr} \\
& + \delpsi{\pbPibf}_{ikr}\pbPi_{ikj} - \delpsi{\pbPibf}_{ikm}\pbPi_{ikm} \delta_{rj}\\
& + \left( D_{\bfJ}^{sym} \psi \right)_{pjk,k} W_{pr} + \left( D_{\bfJ}^{sym} \psi \right)_{pjk} J_{prk} \left. \right\} \left. \right]\, D_{rj} \,dv\\
+ \int_{\partial B(t)}   \rho \left( D_{\bfJ}^{sym} \psi \right)_{pjk} & W_{pr}n_k \, D_{rj} \, da\\
+ \int_{B(t)} \left[ \right. \Lambda_{ik} -  e_{imn}\, \rho \left( \right. & D_{\bfJ}^{sym} \psi \left. \right)_{rnk} W_{rm} \, \left. \right] \, \left( - \frac{1}{2}  e_{ipw} \Omega_{pw,k} \right) \, dv.
\end{split}
\end{equation}
The remaining terms in the dissipation $\textsf{D}$ are
\begin{equation}\label{irrev}
\begin{split}
   - \int_{B(t)}  \left[ \right. & \delpsi{\bfW}_{ij} e_{jrs} \alpha_{ir} + \delpsi{\bfS}_{rwk}  \left( - e_{pjs} W^{-1}_{wl} \alpha_{lj} S_{rpk} \right)  \\
   & - \delpsi{\bfJ}_{rwk,k} e_{wjs} \alpha_{rj} \left. \right] V^{\alpha}_{s}\, dv \\
- \int_{B(t)} \left[ \right. & \delpsi{\bfS}_{rwk}  \left( - e_{kjs} W^{-1}_{wl}\Pi_{rlj} \right)  \\
 & + \delpsi{\pbPibf}_{rwk,m} (\delta_{kp}\delta_{ms} - \delta_{ks}\delta_{mp} ) \Pi_{rwp} \left. \right] V^{\Pi}_s\, dv \\
 + \int_{B(t)} & \delpsi{\bfS}_{rwk,k} W_{wp} S_{rpj} V^S_j \, dv\\
- \int_{\partial B(t)} & \delpsi{\bfS}_{rwk}  n_k W_{wp} S_{rpj} V^{S}_j \, da\\
- \int_{\partial B(t)} & \delpsi{\bfJ}_{rwk}  n_k \alpha_{rj} e_{wjs} V^\alpha_s \, da\\
+ \int_{\partial B(t)} & \delpsi{\pbPibf}_{rwk} n_m (\delta_{kp}\delta_{ms} - \delta_{ks}\delta_{mp} ) \Pi_{rwp} V^{\Pi}_s \, da.
\end{split}
\end{equation}
\subsection{Reversible response and dissipative driving forces}\label{sec_rev_diss}
We deduce ingredients of general constitutive response from the
characterization of the dissipation in Section (\ref{mech_diss}).
\begin{enumerate}
\item It is a physical requirement that the pointwise dissipation
density be invariant under superposed rigid body motions (SRBM) of
the body. The `coefficient' tensor of the spin tensor $\bfOmega$
in the first integrand of (\ref{reversible}) transforms as an
objective tensor under superposed rigid motions (i.e. $ ( \cdot )
\rightarrow \bfQ ( \cdot ) \bfQ^T$ for all proper orthogonal
$\bfQ$), but the spin tensor itself does not (it transforms as
$\bfOmega \rightarrow - \bfomega + \bfQ \bfOmega \bfQ^T$, where
$\bfomega(t) = \dot{\bfQ}(t)\bfQ^T$). Since an \emph{elastic
response} (i.e. $\bfV^\alpha = \bfV^S = \bfV^\Pi = \bf0$) has to
be a special case of our theory and the $2^{nd},3^{rd}$, and
$4^{th}$ integrals of (\ref{reversible}) remain invariant under
SRBM, the coefficient tensor of $\bfOmega$ must vanish. This is a
stringent requirement validating the nonlinear time-dependent
kinematics of the model. Using the Ericksen identity
(\ref{ericksen}), it is verified that the requirement is indeed
satisfied by our model. 
\item We would like to define elastic response as being
non-dissipative, i.e. $\textsf{D} = 0$. Sufficient conditions
ensuring this are given by the following constitutive choices for
$\bfLambda^{dev}$, the \emph{deviatoric part} of the couple stress tensor, the
\emph{symmetric part} of the Cauchy stress tensor, and a boundary
condition:
\begin{equation}\label{couple_stress}
\Lambda_{jk}^{dev} =  e_{jpw} \, \rho \, W_{pr}^T\left( D_{\bfJ}^{sym} \psi \right)_{rwk} ,
\end{equation}
\begin{equation}\label{cauchy_stress}
\begin{split}
T_{rj} + T_{jr} & = \mathcal{A}_{rj} + \mathcal{A}_{jr}\\
\mathcal{A}_{rj} := & \rho \left\{ \right. - \delpsi{\bfW}_{ij} W_{ir}\\
& + \delpsi{\bfS}_{mrk} S_{mjk} - \delpsi{\bfS}_{mwj} S_{mwr} \\
& - \delpsi{\bfJ}_{pjk} J_{prk} - \delpsi{\bfJ}_{mwj} J_{mwr} \\
& + \delpsi{\pbPibf}_{ikr}\pbPi_{ikj} - \delpsi{\pbPibf}_{ikm}\pbPi_{ikm} \delta_{rj}\\
& + \left( D_{\bfJ}^{sym} \psi \right)_{pjk,k} W_{pr} + \left( D_{\bfJ}^{sym} \psi \right)_{pjk} J_{prk} \left. \right\}\\
\end{split}
\end{equation}
and
\begin{equation}\label{sym_hyper_bc}
\begin{split}
\left[ \mathcal{B}_{pwk} + \mathcal{B}_{wpk} \right] n_k & = 0 \ \ \mbox{on boundary of body}\\
\mathcal{B}_{pwk} & := \rho \, W_{pr}^T \left( D_{\bfJ}^{sym} \psi \right)_{rwk}.
\end{split}
\end{equation}

These constitutive choices are meant to be valid for all
processes, whether dissipative or not. The following observations
are in order:
\begin{itemize}
\item The skew-symmetric part of the Cauchy stress, ${\bfT}^{skw}$, is constitutively undetermined (cf.
\cite{mindlin1962effects}). Similarly, the hydrostatic part of the couple stress tensor is constitutively undetermined (cf. \cite{upadhyay2013elastic}), since  $e_{ipw} \Omega_{pw,k} = - \left(e_{iwp} v_{p,w} \right)_{,k}$ in (\ref{reversible}) is deviatoric as the vorticity, being the $curl$ of the velocity field, is necessarily divergence-free. Taking the $curl$ of the balance of angular momentum
(\ref{gov_eqns}$_3$) and substituting the divergence of ${\bfT}^{skw}$ in the balance of the linear momentum
(\ref{gov_eqns}$_2$), one derives a higher order equilibrium
equation between the symmetric part of the Cauchy stress ${\bfT}^{sym}$ and the deviatoric couple-stress ${\bfLambda}^{dev}$:
\begin{equation}\label{higherorderequilibriumequation}
\rho\dot{\bfv} =  \divergence {\bfT}^{sym} + \frac{1}{2} curl(
\divergence {\bfLambda}^{dev}) + \rho\bfb + \frac{1}{2}  curl
\rho\bfK
\end{equation}
In each specific problem, the fields $\rho, \bfx, \bfW, \bfS$ are
obtained by solving (\ref{gov_eqns}$_{1,4,5,6}$) and
(\ref{higherorderequilibriumequation}). This amounts to solving all of (\ref{gov_eqns}), where balance of angular momentum is understood as solved simply by evaluating the skew part of the Cauchy stress from (\ref{gov_eqns}$_3$).
\item The boundary condition (\ref{sym_hyper_bc}) does not
constrain the specification of couple stress related boundary
conditions in any way. 
\item Couple-stresses arise only if the push-forward of the tensor
$D^{sym}_{\bfJ} \psi$ to the current configuration has a
skew-symmetric component. In particular, if $\left( D^{sym}_{\bfJ}
\right)_{rwk} = 0$, then there are no couple-stresses in the model
and, in the absence of body-couples, the stress tensor is
symmetric and balance of linear momentum (\ref{gov_eqns}$_2$),
viewed as the basic equation for solving for the  position field
$\bfx$ or velocity field $\bfv$ is of lower-order (in the sense of
partial differential equations) compared to the situation when
couple-stresses are present. 
\item The important physical case of \emph{dislocation mechanics}
is one where $\left( D^{sym}_{\bfJ} \psi \right)_{rwk} = 0$. Here,
the stored-energy function depends upon $\bfJ = grad\, \bfW$ only
through $\widetilde{\bfalpha} = - \bfJ:\bfX$ and $\delpsi{\bfS} =
\delpsi{\pbPibf} = \bf0$. The theory, including dissipative
effects, then reduces to the one presented in
\cite{acharya2004constitutive, acharya2011microcanonical}.

\item In the \emph{compatible, elastic case}, assuming the
existence of a stress-free reference configuration from which the
deformation is defined with deformation gradient field $\bfF$, we
have $\bfW = \bfF^{-1}$ and the energy function is only a function
of  $grad \, \bfF^{-1}$, and $\bfF^{-1}$. In this case,
$\delpsi{\bfJ}_{pjk} = \left( D^{sym}_{\bfJ} \psi \right)_{pjk}$.
Defining
\[
\psi \left( \bfF^{-1}, grad \, \bfF^{-1} \right) := \tilde{\psi} \left( \bfF \left( \bfF^{-1} \right), Grad\,\bfF \left( \bfF^{-1}, grad \, \bfF^{-1} \right) \right)
\]
and using the relations
\[
\left( Grad \,\bfF \right)_{sP,K} = \left( grad\,\bfF \right)_{sP,k} F_{kK}
\]
\[
\left( grad\,\bfF \right)_{aB,c} = - F_{aM}\left( grad\,\bfF^{-1} \right)_{Mn,c} F_{nB}
\]
along with further manipulation, it can be shown that
\begin{equation}\label{toupin_couple_stress}
\begin{split}
\Lambda_{jk} & = e_{jwp}\, \mathcal{H}_{wpk}\\
\mathcal{H}_{wpk} & = \rho F_{wB}\frac{\partial \tilde{\psi}}{\partial F_{pB,K}}F_{kK}
\end{split}
\end{equation}
and
\begin{equation}\label{toupin_stress}
\mathcal{A}_{rj} \big|_{compatible} = \frac{\partial \tilde{\psi}}{\partial F_{rA}}F_{jA} + \frac{\partial \tilde{\psi}}{\partial F_{rB,C}}F_{jB,C} - \mathcal{H}_{jrk,k}.
\end{equation}
The couple-stress and symmetric part of Cauchy stress  relations that arise from relations (\ref{toupin_couple_stress} - \ref{toupin_stress}) are precisely the ones derived by Toupin \cite{toupin1964theories}\footnote{In the published version of our paper, this reference by Toupin was incorrectly referred to as \cite{toupin1962elastic}.},\cite{truesdell2004non}, starting from a different (static and variational) premise and invoking the notion of an \emph{hyperstress tensor}, a construct we choose not to utilize. Admittedly, we then need a slightly restricted boundary condition (\ref{sym_hyper_bc}), but we do not consider this as a major restriction given the difficulty in physical identification of hyperstresses and hypertractions.
\end{itemize}
\item We refer to dissipative `driving forces' in this context as
the power-conjugate objects to the fields $\bfV^\Pi, \bfV^\alpha,$
and $\bfV^S$ in the dissipation $\textsf{D}$ (\ref{irrev}), since
in their absence there can be no mechanical dissipation in the
theory (i.e. all power supplied to the body is converted in
entirety to stored energy), with the reversible response relations
(\ref{couple_stress}),(\ref{cauchy_stress}),(\ref{sym_hyper_bc})
in effect. Interestingly, the theory suggests separate driving
forces in the bulk and at external boundaries of the body. 
\begin{itemize}
\item The \emph{bulk} driving forces are given by
\begin{equation}\label{bulk_V_Pi}
\begin{split}
V^{\alpha}_{s} \leadsto  - \left[ \right. & \delpsi{\bfW}_{ij} e_{jrs} \alpha_{ir} + \delpsi{\bfS}_{rwk}  \left( - e_{pjs} W^{-1}_{wl} \alpha_{lj} S_{rpk} \right)  \\
   & - \delpsi{\bfJ}_{rwk,k} e_{wjs} \alpha_{rj} \left. \right]
\end{split}
\end{equation}
\begin{equation}\label{bulk_V_Pi}
\begin{split}
V^{\Pi}_s \leadsto - \left[ \right. & \delpsi{\bfS}_{rwk}  \left( - e_{kjs} W^{-1}_{wl}\Pi_{rlj} \right)  \\
 & +  ( \, \delpsi{\pbPibf}_{rwp,s} - \delpsi{\pbPibf}_{rws,p} ) \Pi_{rwp} \left. \right]
\end{split}
\end{equation}
\begin{equation}\label{bulk_V_S}
V^S_j \leadsto \delpsi{\bfS}_{rwk,k} W_{wp} S_{rpj}
\end{equation}
\item The \emph{boundary} driving forces at an external boundary point with outward unit normal $\bfn$ are given by
\begin{equation}\label{boundary_V_S}
V^S_j \leadsto - \delpsi{\bfS}_{rwk}  n_k W_{wp} S_{rpj}
\end{equation}
\begin{equation}\label{boundary_V_alpha}
V^\alpha_s \leadsto - \delpsi{\bfJ}_{rwk}  n_k \alpha_{rj} e_{wjs}
\end{equation}
\begin{equation}\label{boundary_V_Pi}
V^{\Pi}_s \leadsto  \left( \delpsi{\pbPibf}_{rwp} n_s - \delpsi{\pbPibf}_{rws} n_p \right) \Pi_{rwp}.
\end{equation}
\end{itemize}
\end{enumerate}

When the various defect velocities are chosen to be in the
directions of their driving forces, then the mechanical
dissipation in the body is guaranteed to satisfy
\[
\textsf{D} \geq 0,
\]
i.e. the rate of energy supply in the model is never less than the rate of storage of energy.
\subsection{A special constitutive dependence}
There are many situations when the atoms of the as-received body
relieved of applied loads can be re-arranged to form a collection
that is stress-free. An example is that of the as-received body
consisting of a possibly dislocated perfect single crystal. Let us
denote such a stress-free collection of the entire set of atoms in
the body as \textsf{R}. When such an atomic structure is
available, it is often true that, up to boundary-effects, there
are non-trivial homogeneous deformations of the structure that
leave it unchanged (modulo rigid body deformations) and this
provides an energetic constraint on possible atomic motions of the
body. In our modeling, we would like to encapsulate this
structural symmetry-related fact as a constitutive energetic
constraint.

When defects of incompatibility are disallowed (e.g. compatible
phase transformations), then the theory already presented suffices
for modeling, employing multiple well-energy functions in the
deformation gradient from the perfect crystal reference with
second deformation gradient regularization. In the presence of
defects, in particular dislocations, and when the focus is the
modeling of individual dislocations, a constitutive modification
may be required. There exists a gradient flow-based modeling
technique for small deformation analysis called the phase-field
method for dislocations \cite{rodney2003phase, wang2010phase,
denoual2004dynamic} that amalgamates the Ginzburg-Landau paradigm
with Eshelby's \cite{eshelby1957determination} eigenstrain
representation of a dislocation loop; for an approach to coupled
phase-transformations and dislocations at finite deformations
within the same paradigm see \cite{levitas2012advanced}. An
adaptation of those ideas within our framework of unrestricted
material and geometric nonlinearity and conservation-law based
defect dynamics requires, for the representation of physical
concepts like the unstable stacking fault energy density, a
dependence of the stored energy on a measure that reflects
deformation of \textsf{R} to the current atomic configuration.
This measure cannot be defined solely in terms of the i-elastic
1-distortion $\bfW$. The following considerations of this section
provides some physical justification for the adopted definition
(\ref{Ws}) of this measure.

Let us approximate the spatial region occupied by \textsf{R} by a
fixed connected spatial configuration $R$. We consider any atom in
\textsf{R}, say at position $\bfX_{\textsf{R}}$, and consider a
neighborhood of atoms of it.  As the deformation of the body
progresses, we imagine tracking the positions of the atoms of this
neighborhood around $\bfX_{\textsf{R}}$. By approximating the
initial and the image neighborhoods by connected domains, one can
define a deformation between them. We assume that this deformation
is well-approximated by a homogeneous deformation with gradient
$\bfF^{\textsf{s}}(\bfX_{\textsf{R}}, t)$. We assume that by some
well-defined procedure this discrete collection of deformation
gradients at each time (one for each atomic position) can be
extended to a field on the configuration $R$
\footnote{Note that such a tensor field is not $\bfF^p$ of
classical elastoplasticity theory; for instance, its invariance
under superposed rigid body motions of the current configuration
is entirely different from that of $\bfF^p$.}, 
with generic point referred to as $\bfX_R$. Since $R$ and $B(t)$
are both configurations of the body, we can as well view the
motion of the body, say $\bfx_R$, with $R$ as a reference
configuration and with deformation gradient field
\[
\bfF_R = Grad_{\bfX_{R}} \bfx,
\]
where the expression on the right hand side refers to the gradient of the position field $\bfx$ on the configuration $R$.

Through this one-to-one motion referred to $R$ we associate the
field
\[
\bfW^{\textsf{s}}(\bfx,t) := \bfF^{\textsf{s}-1}(\bfx,t)
\]
with the current configuration $B(t)$ in the natural way and constrain the possible local deformations $\bfF^{\textsf{s}}$
\footnote{This may also be viewed as a constraint on the atomic re-arrangement leading to the choice of the particular \textsf{R}.}
 by requiring
\[
curl \, \bfW^{\textsf{s}} = \widetilde{\bfalpha} \implies curl \, (\bfW - \bfW^{\textsf{s}} ) = \bf0
\]
and choosing the `free' gradient of a vector field through
\begin{equation}\label{Ws}
\bfW = \bfW^{\textsf{s}} + grad\, \bfx^{-1}_R \implies \bfW^{\textsf{s}} := \bfW - \bfF^{-1}_R.
\end{equation}
We note that the knowledge of the motion of the body and the
evolution of the $\bfW$ field completely determine the evolution
of the field $\bfW^{\textsf{s}}$. In the manner defined, in
principle $\bfW^{\textsf{s}}$ is an unambiguously initializable
field whenever the atomic configuration in the as-received body is
known and a `perfect' atomic structure \textsf{R} for the body
exists.

When a dependence of the energy function on the structural
distortion is envisaged, this implies an additional dependence of
the stored energy function (\ref{free_energy}) on $\bfF_R$ (and a
dependence on the configuration \textsf{R}). This implies
corresponding changes in the Ericksen identity, reversible
response functions, and the driving forces that may be deduced
without difficulty.

We emphasize, however, that it is not clear to us at this point
that the constitutive modeling necessarily requires accounting for
the structural variable $\bfW^{\textsf{s}}$ (or equivalently the
pair $\bfW$ and $\bfF_R$), despite the viewpoint of the
phase-field methodology. In particular, whether a suitable
dependence of the stored energy function solely on the element
$\bfW$ of the pair suffices for the prediction of observed
behavior related to motion of individual dislocations needs to be
explored in detail.

\section{`Small deformation' model}
In this section we present a model where many of the geometric
nonlinearities that appear in the theory presented in Section
\ref{MFI_thermo} are ignored. This may be considered as an
extension of the theory of  \emph{linear elasticity} to account
for the dynamics of phase boundaries, g.disclinations, and
dislocations. A main assumption is that the all equations are
posed on a fixed, known, configuration that enters
`parametrically' in the solution to the equations. Such a model
has been described in \cite{acharya2012coupled}. In what we
present here, there is a difference in the reversible responses
from those proposed in \cite{acharya2012coupled}, even though the
latter also ensure that the dissipation vanishes in the model for
elastic processes. The choices made here render our model
consistent with Toupin's \cite{toupin1962elastic} model of
higher-order elasticity in the completely compatible case.

The eigenwall field in the small deformation case is denoted by
$\hat{\bfS}$. All g.disclination density measures are denoted by
$\hat{\bfPi}$. The elastic 1-distortion is approximated by $\bfI -
\bfU^e$ where $\bfU^e$ is a `small' elastic distortion measure and
we further introduce a \emph{plastic distortion} field by the
definition
\[
\bfU^e := grad\, \bfu - \bfU^p,
\]
where $\bfu$ is the displacement field of the body from the given
distinguished reference configuration. The strain tensor is
defined as $\bfveps : = \left( grad\, \bfu \right)_{sym}$. The
elastic 2-distortion is defined as $\bfG^e := grad \, \bfU^e +
\hat{\bfS}$, with the g.disclination density  as $\curl \, \bfG^e
= curl \, \hat{\bfS} = \hat{\bfPi}$. The dislocation density is
defined as $\hat{\bfalpha} := - \bfG^e:\bfX = curl\, \bfU^e -
\hat{\bfS}:\bfX$.

The governing equations are
\begin{equation}\label{small_def_gov_eqns}
\begin{split}
\rho \ddot{\bfu} & = div\, \bfT + \hat{\bfb}\\
\bf0 & = div\, \bfLambda  - \bfX : \bfT + \hat{\bfK} \\
\dot{\bfU}^p & = \hat{\bfalpha} \times \hat{\bfV}^\alpha \\
\dot{\hat{\bfS}} & = - \bfPi \times \hat{\bfV}^\Pi + grad \left( \hat{\bfS}\hat{\bfV}^S \right) \\
\dot{\hat{\bfPi}} & = - curl \left( \hat{\bfPi} \times \hat{\bfV}^\Pi \right).
\end{split}
\end{equation}
Here $\hat{\bfV}^S$ is the eigenwall velocity, $\hat{\bfV}^\alpha$
the dislocation velocity, $\hat{\bfV}^\Pi$ the disclination
velocity, and $\hat{\bfb}$ and $\hat{\bfK}$ are body force and
couple densities per unit volume. We also define $\hat{\bfJ} :=
grad \, \bfU^e$.

The stored energy density response (per unit volume of the
reference) is assumed to have the following dependencies:
\[
\psi = \psi \left( \bfU^e, \hat{\bfS}, \hat{\bfPi}, \hat{\bfJ} \right),
\]
and a necessary condition for the invariance of the energy under
superposed infinitesimal rigid deformations is
\[
\delpsi{\bfU^e}:\bfs = 0 \ \ \mbox{for all skew tensors} \ \bfs,
\]
which implies that $\delpsi{\bfU^e}$ has to be a symmetric tensor,
thus constraining the functional form of $\psi$.

On defining $\left( D^{sym}_{\hat{\bfJ}} \psi \right)_{ijk} := \frac{1}{2} \left[ \delpsi{\hat{\bfJ}}_{ijk} + \delpsi{\hat{\bfJ}}_{ikj} \right]$, the dissipation can be characterized as:
\begin{equation}\label{small_def_diss}
\begin{split}
\textsf{D}  = & \int_ B T_{ij} \dot{\eps}_{ij}\, dv - \frac{1}{2} \int_ B \Lambda_{ij} e_{irs} \Omega_{rs,j} \, dv - \int_ B \dot{\psi} \, dv \\
  = & \int_ B \left[ T_{ij} - \delpsi{\bfU^e}_{ij} + \left( D^{sym}_{\hat{\bfJ}} \psi \right)_{ijk,k} \right] \, D_{ij} \, dv \\
& + \int_ B \left[ - \frac{1}{2} e_{irs} \Lambda_{ij} - \left( D^{sym}_{\hat{\bfJ}} \psi \right)_{rsj} \right] \, \Omega_{rs,j} \, dv \\
& + \int_ B \left[ e_{sjr} \left\{ \delpsi{\bfU^e}_{ij} - \delpsi{\hat{\bfJ}}_{ijk,k} \right\} \right] \hat{\alpha}_{ir} \hat{V}^\alpha_s \, dv \\
& + \int_ B \delpsi{\hat{\bfS}}_{ijk,k}\hat{S}_{ijr} \hat{V}^S_r \, dv\\
& + \int_ B \left[ e_{snr} \left\{ \delpsi{\hat{\bfS}}_{ijn} + e_{nmk} \delpsi{\hat{\bfPi}}_{ijk,m} \right\}  \right] \hat{\Pi}_{ijr}\, \hat{V}^\Pi_s \, dv \\
& - \int_{\partial B} \left( D^{sym}_{\hat{\bfJ}} \psi \right)_{ijk} n_k \dot{\bfveps}_{ij} \, da \\
& + \int_{\partial B} e_{sjr} \delpsi{\hat{\bfJ}}_{ijk} n_k \hat{\alpha}_{ir} \hat{V}^\alpha_s \, da \\
& - \int_{\partial B} \delpsi{\hat{\bfS}}_{ijk} n_k \hat{S}_{ijr} \hat{V}^S_r \, da\\
& + \int_{\partial B} \delpsi{\hat{\bfPi}}_{ijk} \left[ \delta_{kr} \delta_{ms} - \delta_{ks}\delta_{mr} \right] n_m \hat{\Pi}_{ijr} \hat{V}^{\Pi}_s \, da.
 \end{split}
\end{equation}

\subsection{Reversible response and driving forces in the small deformation model}
Motivated by the characterization (\ref{small_def_diss}), we
propose the following constitutive guidelines that ensure
non-negative dissipation in general and vanishing dissipation in
the elastic case:
\begin{equation}\label{small_def_const}
\begin{split}
T_{ij} + T_{ji} & = \hat{\cal{A}}_{ij} +  \hat{\cal{A}}_{ji} \\
\hat{\cal{A}}_{ij} & := \delpsi{\bfU^e}_{ij} - \left( D^{sym}_{\hat{\bfJ}} \psi \right)_{ijk,k}\\
\Lambda_{ij}^{dev} & = - e_{irs} \left( D^{sym}_{\hat{\bfJ}} \psi \right)_{rsj} \\
& \left[ \left( D^{sym}_{\hat{\bfJ}} \psi \right)_{ijk}  + \left( D^{sym}_{\hat{\bfJ}} \psi \right)_{jik} \right] n_k \Big|_{boundary} = 0 \\
\hat{V}^\alpha_s \big|_{bulk} & \leadsto  e_{sjr} \left\{ \delpsi{\bfU^e}_{ij} - \delpsi{\hat{\bfJ}}_{ijk,k} \right\}  \hat{\alpha}_{ir}\\
\hat{V}^S_r \big|_{bulk} & \leadsto \delpsi{\hat{\bfS}}_{ijk,k}\hat{S}_{ijr} \\
\hat{V}^\Pi_s \big|_{bulk} & \leadsto e_{snr} \left\{ \delpsi{\hat{\bfS}}_{ijn} + e_{nmk} \delpsi{\hat{\bfPi}}_{ijk,m} \right\} \hat{\Pi}_{ijr}\\
\hat{V}^\alpha_s \big|_{boundary} & \leadsto  e_{sjr} \delpsi{\hat{\bfJ}}_{ijk} n_k \, \hat{\alpha}_{ir}\\
\hat{V}^S_r \big|_{boundary} & \leadsto - \delpsi{\hat{\bfS}}_{ijk} n_k \, \hat{S}_{ijr} \\
\hat{V}^{\Pi}_s \big|_{boundary} & \leadsto  \left[ \delpsi{\hat{\bfPi}}_{ijr} n_s - \delpsi{\hat{\bfPi}}_{ijs} n_r\right] \hat{\Pi}_{ijr}.
\end{split}
\end{equation}

As before, a dependence of the energy on $\bfF^{\textsf{s}}$ in the nonlinear case translates to an extra dependence of the stored energy on
\[
\bfU^p = grad\, \bfu - \bfU^e = \bfI - \bfU^e - \left (\bfI - grad\, \bfu \right) \approx \bfW - \bfF_R^{-1} = \bfW^{\textsf{s}}
\]
in the small deformation case, with corresponding changes in the reversible response and driving forces.
\section{Contact with the differential geometric point of view}\label{diff-geom}
For the purpose of this section it is assumed that we operate on a
simply-connected subset of the current configuration $B$.
Arbitrary (3-d) curvilinear coordinate systems for the set will be
invoked as needed, with the generic point denoted as $\left(\xi^1,
\xi^2, \xi^3 \right)$. Lower-case Greek letters will be used to
denote indices for such coordinates. The natural basis of the
coordinate system on the configuration $B$ will be denoted as the
list of vectors
\[
\bfe_\alpha = \frac{\partial \bfx}{\partial \xi^\alpha} \ \ \alpha = 1,2,3,
\]
with dual basis $\left( \bfe^\beta = grad\,\xi^\beta, \beta =
1,2,3 \right)$. We will assume all fields to be as smooth as
required; in particular, equality of second partial derivatives
will be assumed throughout.

Beyond the physical motivation provided for it in Section
\ref{g_disc_ewall} as a line density carrying a tensorial
attribute, the disclination density field $\bfPi = curl \, \bfY $
alternatively characterizes whether a solution $\tilde{\bfW}$
($2^{nd}$-order tensor field) exists to the equation
\begin{equation}\label{cont_mech_int}
grad\,\tilde{\bfW} = \bfY,
\end{equation}
with existence guaranteed when $\bfPi = curl \, \bfY = curl \,
\bfS = \bf0$ which, in a rectangular Cartesian coordinate system,
amounts to
\begin{equation}\label{cont_mech_int_cond}
S_{ijk,l} - S_{ijl,k} = e_{rlk}e_{rqp}S_{ijp,q} = e_{rlk} \left( curl \, \bfS \right)_{ijr} = 0.
\end{equation}

This is a physically meaningful question in continuum mechanics
with a simple answer. Moreover, when such a solution exists, the
existence of a triad $ \tilde{\bfd}_\alpha, \alpha = 1,2,3\ $ of
vectors corresponding to each choice of a coordinate system for
$B$ is also guaranteed by the definition
\[
\tilde{\bfd}_{\alpha} := \tilde{\bfW}\bfe_\alpha.
\]
This question of the existence of a triad of vectors related to
arbitrary coordinate systems for $B$ and the integrability of
$\bfY$ can also be posed in a differential geometric context,
albeit far more complicated.

We first consider the i-elastic 1-distortion $\bfW$ that is
assumed to be an \emph{invertible} $2^{nd}$-order tensor field by
definition. Defining
\[
\bar{\bfd}_\alpha = \bfW \bfe_\alpha
\]
and noting that $\bar{\bfd}_\alpha, \alpha = 1,2,3$ is necessarily
a basis field, there exists an array $\bar{\Gamma}^\mu_{\alpha
\beta}$ satisfying
\begin{equation}\label{d_bar}
\bar{\bfd}_{\alpha,\beta} = \bar{\Gamma}^\mu_{\alpha \beta} \bar{\bfd}_\mu.
\end{equation}
Let the dual basis of $\left( \bar{\bfd}_\alpha, \alpha = 1,2,3 \right)$ be $\left( \bar{\bfd}^\alpha = \bfW^{-T} \bfe^\alpha, \alpha = 1,2,3 \right)$. Then
\[
\bar{\Gamma}^{\rho}_{\alpha \beta} = \bfe^\rho \cdot \bfW^{-1} \left( \left[ \left\{ grad\, \bfW \right\} \bfe_\beta \right] \bfe_\alpha + \bfW \bfe_{\alpha,\beta} \right).
\]
We observe that even though (\ref{d_bar}) is an overconstrained
system of $9$ vector equations for $3$ vector fields, solutions
exist due to the invertibility of $\bfW$, and the following
`integrability' condition arising from $\bar{\bfd}_{\alpha, \beta
\gamma} = \bar{\bfd}_{\alpha, \gamma \beta}$, holds:
\begin{equation}\label{gamma_bar_int}
\bar{\Gamma}_{\alpha \beta, \gamma}^\mu - \bar{\Gamma}_{\alpha \gamma, \beta}^\mu +  \bar{\Gamma}^\rho_{\alpha \beta} \bar{\Gamma}^\mu_{\rho \gamma} - \bar{\Gamma}^\rho_{\alpha \beta} \bar{\Gamma}^\mu_{\rho \gamma}  = 0.
\end{equation}

Guided by the integrability/existence question suggested by
(\ref{d_bar}) we now turn the argument around and ask for
conditions of existence of a vector field triad $\left(
\bfd_\alpha \right)$ \emph{given} the connection symbols $\Gamma$
defined by 
\begin{equation*}
\begin{split}
\Gamma^\rho_{\alpha \beta} & := \bar{\Gamma}^\rho_{\alpha \beta} + S^\rho_{. \alpha \beta} \\
S^\rho_{. \alpha \beta} & := \bfe^\rho \cdot \bfW^{-1} \left( \left\{ \bfS \bfe_\beta \right\} \bfe_\alpha \right).
\end{split}
\end{equation*}
Thus, we ask the question of existence of smooth solutions to
\begin{equation}\label{diff_geom_int}
\bfd_{\alpha,\beta} = \Gamma^\mu_{\alpha \beta} \bfd_\mu.
\end{equation}
Based on a theorem of Thomas \cite{thomas1934systems}, it can be
shown that a $9$-parameter family of (global) solutions on
simply-connected domains may be constructed when the following
condition on the array $\Gamma$ holds:
\begin{equation}\label{diff_geom_int_cond}
R^\mu_{.\alpha \beta \gamma}(\Gamma):= \Gamma_{\alpha \beta, \gamma}^\mu - \Gamma_{\alpha \gamma, \beta}^\mu +  \Gamma^\rho_{\alpha \beta} \Gamma^\mu_{\rho \gamma} - \Gamma^\rho_{\alpha \beta} \Gamma^\mu_{\rho \gamma}
 = 0.
\end{equation}
The condition corresponds to the mixed components of the curvature
tensor of the connection $\Gamma$ vanishing and results in
$\bfd_{\alpha, \beta \gamma} = \bfd_{\alpha, \gamma \beta}$ for
the $\left( \bfd_\alpha \right)$ triad that can be constructed. We
note that
\[
R^\alpha_{. \mu \beta \gamma} (\Gamma) = R^\alpha_{. \mu \beta \gamma} (\bar{\Gamma}) + R^\alpha_{. \mu \beta \gamma} (S) + \bar{\Gamma}^\alpha_{\nu \gamma} S^\nu_{. \mu \beta} + \bar{\Gamma}^\nu_{\mu \beta} S^\alpha_{. \nu \gamma} - \bar{\Gamma}^\alpha_{\nu \beta} S^\nu_{. \mu \gamma} - \bar{\Gamma}^\nu_{\mu \gamma} S^\alpha_{. \nu \beta}
\]
with $R^\alpha_{. \mu \beta \gamma} (\bar{\Gamma}) = 0$ from
(\ref{gamma_bar_int}). Furthermore, the typical differential
geometric treatment \cite{kondo1955non, bilby1960continuous,
kroner1992gauge, clayton2006modeling}  imposes the condition of a
metric differential geometry, i.e. the covariant derivative of the
metric tensor (here $\bfW^T\bfW$) with respect to the connection
$\Gamma$ is required to vanish. There is no need in our
development to impose any such requirement.

The difference in complexity of the continuum mechanical and
differential geometric integrability conditions
(\ref{cont_mech_int_cond}) and (\ref{diff_geom_int_cond}), even
when both are expressed in rectangular Cartesian coordinates, is
striking. It arises because of the nature of the existence
questions asked in the two cases. The differential geometric
question (\ref{diff_geom_int}) involves the unknown vector field
on the right hand side while the continuum mechanical question
(\ref{cont_mech_int}), physically self-contained and sufficiently
general for the purpose at hand, is essentially the question from
elementary vector analysis of when a potential exists for a
\emph{completely prescribed} vector field.

Finally, we note that both in the traditional metric differential
geometric treatment of defects \cite{kondo1955non,
bilby1960continuous, kroner1992gauge, clayton2006modeling} and our
continuum mechanical treatment at finite strains, it is not
straightforward, if possible at all, to separate out the effects
of strictly rotation-gradient and strain-gradient related
incompatibilities/non-integrabilities. Fortunately from our point
of view, this is not physically required either (for specifying,
e.g., the defect content of a terminating elastic distortion
discontinuity from observations).


\section{Concluding remarks}
A new theoretical approach for studying the coupled dynamics of
phase transformations and plasticity in solids has been presented.
It extends nonlinear elasticity by considering new continuum
fields arising from defects in compatibility of deformation. The
generalized eigendeformation based kinematics allows a natural
framework for posing \emph{kinetic} balance/conservation laws for
defect densities and consequent dissipation, an avenue not
available through simply higher-gradient, `capillary'/surface
energy regularizations of compatible theory. Such a feature is in
the direction of theoretical requirements suggested by results of
sharp-interface models from nonlinear elasticity in the case of
phase transformations \cite{abeyaratne2006evolution}. In addition, finite-total-energy,
non-singular, defect-like fields can be described (that may also
be expected to be possible with higher-gradient regularizations),
and their evolution can be followed without the cumbersome tracking
of complicated, evolving, multiply-connected geometries. This feature has obvious beneficial implications for practical numerical implementations where the developed model introduces interesting combinations of elliptic and hyperbolic (when material inertia is included) systems with degenerate parabolic equations for numerical discretization. The elliptic component includes $div-curl$ systems, novel in the context of their use in solid mechanics. Significant components of such problems have been dealt with computationally in our prior work e.g. \cite{roy2005finite,varadhan2006dislocation,fressengeas2011elasto,taupin_shear_coupled}, and detailed considerations for the present model will be the subject of future work.

The generalized eigendeformation fields have striking similarities
with gauge fields of high-energy particle physics,  but do not
arise from considerations of gauge invariance of an underlying
Hamiltonian. Instead, they arise from the physical requirement of
modeling finite total energies in bodies that contain commonly
observed 1 and 2-dimensional defects, and from a desire to be able
to model their observed motion and interactions.

In formulating a continuum mechanical model of solid-solid phase
transformation behavior based squarely on the kinematics of
deformation incompatibility, our work differs from that of
\cite{fried1994dynamic} and those of \cite{khachaturian1983theory,
roitburd1978martensitic}. In the context of dislocation plasticity
alone, for the same reason it differs from the strain-gradient
plasticity work of \cite{aifantis1984microstructural,
fleck2001reformulation, gao1999mechanism}. There is an extended
body of work in strain-gradient plasticity that accounts for the
dislocation density in some form \cite{steinmann1996views,
gurtin2002gradient, forest2003gradient, evers2004non,
levkovitch2006large, kuroda2008finite, gudmundson2004unified,
fleck2009mathematical} but none have been shown to build up from a
treatment of the statics and dynamics of individual dislocations
as in our case \cite{acharya2001model, acharya2003driving,
varadhan2006dislocation, das2012can, zhu-chapman,
taupin_shear_coupled}.

Finally, we mention a widely used, and quite successful, framework
for grain-boundary network evolution \cite{mullins1956two,
kinderlehrer2006variational, elsey2009diffusion}. This involves
postulating a grain boundary energy density based on
misorientation and the normal vector to the boundary and evolving
the network based on a gradient flow of this energy (taking
account of the natural boundary condition that arises at triple
lines). Given that a grain boundary is after all a sharp
transition layer in lattice orientation and the latter is a part
of the elastic distortion of a lattice that stretches and bends to
transmit stresses and moments, it is reasonable to ask why such
modeling succeeds with the complete neglect of any notions of
stress or elastic deformation and what the model's relation might
be to a theory where stresses and elastic strains are not
constrained to vanish. The Mullins model does not allow asking
such questions. With localized concentrations of the eigenwall
field representing the geometry of grain boundaries (including
their normals), g.disclinations representing triple (or higher)
lines, dependence of the energy on the eigenwall field and the
i-elastic 1-distortion representing effects of misorientation, and
the eigenwall velocity representing the grain boundary velocity,
our model provides a natural framework, accounting for
compatibility conditions akin to Herring's relation at triple
lines, for the response of grain boundaries to applied stress
\cite{taupin_shear_coupled,fressengeas2012disclination}. Moreover,
it allows asking the question of whether stress-free
initializations can remain (almost) stress-free on evolution.
Interestingly, it appears that it may be possible to even have an
exact analog of the stress-free/negligible stress model by
allowing for general evolution of the eigenwall field $\bfS$, and
constraining the dislocation density field $\bfalpha$ to ensure
that $\widetilde{\bfalpha} = - curl \, \bfW$ always belongs to the
space of $curls$ of (proper-orthogonal tensor) rotation fields. We
leave such interesting physical questions for further study along
with the analysis of `simple' ansatz-based, exact reduced models of phase
boundary evolution coupled to dislocation plasticity within our setting that have been formulated.

Ericksen \cite{ericksen1998nonlinear, ericksen2008cauchy} raises interesting and important questions about the (in)adequacy of modeling crystal defects with nonlinear elasticity, the interrelationships between the mechanics of twinning and dislocations, and the conceptual (un)importance of involving a reference configuration in the mechanics of crystalline solids, among others. It is our hope that we have made a first step in answering such questions with the theory presented in this paper.

\section*{Acknowledgments}
We thank Rohan Abeyaratne and David Parks for their comments on this paper. AA acknowledges the hospitality and support of the Laboratoire d'Etude des Microstructures et de M\'ecanique des Mat\'eriaux (LEM3), Universit\'e de Lorraine/CNRS, during visits in the summers of 2010 and 2011 when this work was initiated. CF acknowledges the hospitality of the Department of Civil \& Environmental Engineering at Carnegie Mellon during visits in 2009 and 2010.
\bibliography{defects}

\newcommand{\etalchar}[1]{$^{#1}$}
\begin{thebibliography}{ZLJVV{\etalchar{+}}04}

\bibitem[ACF99]{anderson1999continuum}
David~R. Anderson, Donald~E. Carlson, and Eliot Fried.
\newblock A continuum-mechanical theory for nematic elastomers.
\newblock {\em Journal of Elasticity}, 56(1):33--58, 1999.

\bibitem[Ach01]{acharya2001model}
Amit Acharya.
\newblock A model of crystal plasticity based on the theory of continuously
  distributed dislocations.
\newblock {\em Journal of the Mechanics and Physics of Solids}, 49(4):761--784,
  2001.

\bibitem[Ach03]{acharya2003driving}
Amit Acharya.
\newblock Driving forces and boundary conditions in continuum dislocation
  mechanics.
\newblock {\em Proceedings of the Royal Society of London. Series A:
  Mathematical, Physical and Engineering Sciences}, 459(2034):1343--1363, 2003.

\bibitem[Ach04]{acharya2004constitutive}
Amit Acharya.
\newblock Constitutive analysis of finite deformation {F}ield {D}islocation
  {M}echanics.
\newblock {\em Journal of the Mechanics and Physics of Solids}, 52(2):301--316,
  2004.

\bibitem[Ach11]{acharya2011microcanonical}
A.~Acharya.
\newblock Microcanonical entropy and mesoscale dislocation mechanics and
  plasticity.
\newblock {\em Journal of Elasticity}, 104:23--44, 2011.

\bibitem[AD12]{acharya-dayal}
A.~Acharya and K.~Dayal.
\newblock {Continuum mechanics of line defects in liquid crystals and liquid
  crystal elastomers}.
\newblock {\em in press Quarterly of Applied Mathematics}, 2012.

\bibitem[AF12]{acharya2012coupled}
Amit Acharya and Claude Fressengeas.
\newblock Coupled phase transformations and plasticity as a field theory of
  deformation incompatibility.
\newblock {\em International Journal of Fracture}, 174(1):87--94, 2012.

\bibitem[Aif84]{aifantis1984microstructural}
E.~C. Aifantis.
\newblock On the microstructural origin of certain inelastic models.
\newblock {\em Journal of Engineering Materials and Technology},
  106(4):326--330, 1984.

\bibitem[AK90]{abeyaratne1990driving}
Rohan Abeyaratne and James~K. Knowles.
\newblock On the driving traction acting on a surface of strain discontinuity
  in a continuum.
\newblock {\em Journal of the Mechanics and Physics of Solids}, 38(3):345--360,
  1990.

\bibitem[AK91]{abeyaratne1991implications}
Rohan Abeyaratne and James~K. Knowles.
\newblock Implications of viscosity and strain-gradient effects for the
  kinetics of propagating phase boundaries in solids.
\newblock {\em SIAM Journal on Applied Mathematics}, 51(5):1205--1221, 1991.

\bibitem[AK06]{abeyaratne2006evolution}
Rohan Abeyaratne and James~K. Knowles.
\newblock {\em Evolution of phase transitions: a continuum theory}.
\newblock Cambridge University Press, 2006.

\bibitem[AZ14]{acharya_zhang}
Amit Acharya and Xiaohan Zhang.
\newblock From dislocation motion to an additive velocity gradient
  decomposition, and some simple models of dislocation dynamics.
\newblock In Ph.~G. Ciarlet and Ta-Tsien Li, editors, {\em Proceedings of the
  International Conference on Nonlinear and Multiscale Partial Differential
  Equations: Theory, Numerics and Applications held at Fudan University,
  Shanghai, September 16--20, 2013 in honor of {L}uc Tartar, Series in
  Contemporary Applied Mathematics}. Higher Education Press (Beijing), and
  World Scientific (Singapore), 2014.

\bibitem[Bil60]{bilby1960continuous}
B.~A. Bilby.
\newblock Continuous distributions of dislocations.
\newblock {\em Progress in Solid Mechanics}, 1(1):329--398, 1960.

\bibitem[BJ87]{ball1989fine}
John~M. Ball and Richard~D. James.
\newblock Fine phase mixtures as minimizers of energy.
\newblock {\em Archive for Rational Mechanics and Analysis}, 100:13--52, 1987.

\bibitem[BK84]{barsch1984twin}
G.~R. Barsch and J.~A. Krumhansl.
\newblock Twin boundaries in ferroelastic media without interface dislocations.
\newblock {\em Physical Review Letters}, 53:1069--1072, 1984.

\bibitem[BZB{\etalchar{+}}12]{bieler2012role}
Thomas~R. Bieler, B.~Zhou, Lauren Blair, Amir Zamiri, Payam Darbandi, Farhang
  Pourboghrat, Tae-Kyu Lee, and Kuo-Chuan Liu.
\newblock The role of elastic and plastic anisotropy of {S}n in
  recrystallization and damage evolution during thermal cycling in {SAC}305
  solder joints.
\newblock {\em Journal of Electronic Materials}, 41(2):283--301, 2012.

\bibitem[Cas04]{casey2004volterra}
James Casey.
\newblock On {V}olterra dislocations of finitely deforming continua.
\newblock {\em Mathematics and Mechanics of Solids}, 9(5):473--492, 2004.

\bibitem[CCF{\etalchar{+}}06]{cui2006combinatorial}
Jun Cui, Yong~S. Chu, Olugbenga~O. Famodu, Yasubumi Furuya, Jae
  Hattrick-Simpers, Richard~D. James, Alfred Ludwig, Sigurd Thienhaus, Manfred
  Wuttig, Zhiyong Zhang, and Ichiro Takeuchi.
\newblock Combinatorial search of thermoelastic shape-memory alloys with
  extremely small hysteresis width.
\newblock {\em Nature Materials}, 5(4):286--290, 2006.

\bibitem[CMB06]{clayton2006modeling}
John~D. Clayton, David~L. McDowell, and Douglas~J. Bammann.
\newblock Modeling dislocations and disclinations with finite micropolar
  elastoplasticity.
\newblock {\em International Journal of Plasticity}, 22(2):210--256, 2006.

\bibitem[DAZM]{das2012can}
Amit Das, Amit Acharya, Johannes Zimmer, and Karsten Matthies.
\newblock Can equations of equilibrium predict all physical equilibria? {A}
  case study from {F}ield {D}islocation {M}echanics.
\newblock {\em Mathematics and Mechanics of Solids}, 18(8):803--822.

\bibitem[Den04]{denoual2004dynamic}
Christophe Denoual.
\newblock Dynamic dislocation modeling by combining {P}eierls {N}abarro and
  {G}alerkin methods.
\newblock {\em Physical Review B}, 70(2):024106, 2004.

\bibitem[deW70]{de1970linear}
R.~deWit.
\newblock Linear theory of static disclinations.
\newblock {\em Fundamental Aspects of Dislocation Theory}, 1:651--673, 1970.

\bibitem[deW73]{dewit_disclinations_II}
R.~deWit.
\newblock Theory of disclinations:{II}. {C}ontinuous and discrete disclinations
  in anisotropic {E}lasticity.
\newblock {\em Journal of Research of the National Bureau of Standards - A.
  Physics and Chemistry}, 77A(1):49--100, 1973.

\bibitem[DW72]{de1972partial}
R~De~Wit.
\newblock Partial disclinations.
\newblock {\em Journal of Physics C: Solid State Physics}, 5(5):529, 1972.

\bibitem[DZ11]{derezin2011disclinations}
S~Derezin and L~Zubov.
\newblock Disclinations in nonlinear elasticity.
\newblock {\em ZAMM-Journal of Applied Mathematics and Mechanics/Zeitschrift
  f{\"u}r Angewandte Mathematik und Mechanik}, 91(6):433--442, 2011.

\bibitem[EBG04]{evers2004non}
L.~P. Evers, W.~A.~M. Brekelmans, and M.~G.~D. Geers.
\newblock Non-local crystal plasticity model with intrinsic {SSD} and {GND}
  effects.
\newblock {\em Journal of the Mechanics and Physics of Solids},
  52(10):2379--2401, 2004.

\bibitem[EES09]{elsey2009diffusion}
Matt Elsey, Selim Esedoglu, and Peter Smereka.
\newblock Diffusion generated motion for grain growth in two and three
  dimensions.
\newblock {\em Journal of Computational Physics}, 228(21):8015--8033, 2009.

\bibitem[Eri98]{ericksen1998nonlinear}
J.~L. Ericksen.
\newblock On nonlinear elasticity theory for crystal defects.
\newblock {\em International Journal of Plasticity}, 14(1):9--24, 1998.

\bibitem[Eri08]{ericksen2008cauchy}
J.~L. Ericksen.
\newblock On the {C}auchy {B}orn {R}ule.
\newblock {\em Mathematics and Mechanics of Solids}, 13(3-4):199--220, 2008.

\bibitem[Esh57]{eshelby1957determination}
J.~D. Eshelby.
\newblock The determination of the elastic field of an ellipsoidal inclusion,
  and related problems.
\newblock {\em Proceedings of the Royal Society of London. Series A.
  Mathematical and Physical Sciences}, 241(1226):376--396, 1957.

\bibitem[FG94]{fried1994dynamic}
Eliot Fried and Morton~E Gurtin.
\newblock Dynamic solid-solid transitions with phase characterized by an order
  parameter.
\newblock {\em Physica D: Nonlinear Phenomena}, 72(4):287--308, 1994.

\bibitem[FH01]{fleck2001reformulation}
N.~A. Fleck and J.~W. Hutchinson.
\newblock A reformulation of strain gradient plasticity.
\newblock {\em Journal of the Mechanics and Physics of Solids},
  49(10):2245--2271, 2001.

\bibitem[FS03]{forest2003gradient}
S.~Forest and R.~Sievert.
\newblock Elastoviscoplastic constitutive frameworks for generalized continua.
\newblock {\em Acta Mechanica}, 160(1-2):71--111, 2003.

\bibitem[FTC11]{fressengeas2011elasto}
C.~Fressengeas, V.~Taupin, and L.~Capolungo.
\newblock An elasto-plastic theory of dislocation and disclination fields.
\newblock {\em International Journal of Solids and Structures},
  48(25):3499--3509, 2011.

\bibitem[FTUC12]{fressengeas2012disclination}
Claude Fressengeas, Vincent Taupin, Manas Upadhyay, and Laurent Capolungo.
\newblock Tangential continuity of elastic/plastic curvature and strain at
  interfaces.
\newblock {\em International Journal of Solids and Structures}, 49:2660--2667,
  2012.

\bibitem[FW09]{fleck2009mathematical}
N.~A. Fleck and J.~R. Willis.
\newblock A mathematical basis for strain-gradient plasticity theory—part i:
  Scalar plastic multiplier.
\newblock {\em Journal of the Mechanics and Physics of Solids}, 57(1):161--177,
  2009.

\bibitem[GHNH99]{gao1999mechanism}
H.~Gao, Y.~Huang, W.~D. Nix, and J.~W. Hutchinson.
\newblock Mechanism-based strain gradient plasticity {I}. {T}heory.
\newblock {\em Journal of the Mechanics and Physics of Solids},
  47(6):1239--1263, 1999.

\bibitem[GLS10]{groger2010incompatibility}
R~Gr{\"o}ger, T~Lookman, and A~Saxena.
\newblock Incompatibility of strains and its application to mesoscopic studies
  of plasticity.
\newblock {\em Physical Review B}, 82(14):144104, 2010.

\bibitem[Gud04]{gudmundson2004unified}
Peter Gudmundson.
\newblock A unified treatment of strain gradient plasticity.
\newblock {\em Journal of the Mechanics and Physics of Solids},
  52(6):1379--1406, 2004.

\bibitem[Gur02]{gurtin2002gradient}
Morton~E Gurtin.
\newblock A gradient theory of single-crystal viscoplasticity that accounts for
  geometrically necessary dislocations.
\newblock {\em Journal of the Mechanics and Physics of Solids}, 50(1):5--32,
  2002.

\bibitem[HH75]{hill1975bifurcation}
R.~Hill and J.~W. Hutchinson.
\newblock Bifurcation phenomena in the plane tension test.
\newblock {\em Journal of the Mechanics and Physics of Solids}, 23(4):239--264,
  1975.

\bibitem[HLL{\etalchar{+}}12]{hefferan2012observation}
Christopher~M Hefferan, Jonathan Lind, Shiu~Fai Li, Ulrich Lienert, Anthony~D
  Rollett, and Robert~M Suter.
\newblock Observation of recovery and recrystallization in high-purity aluminum
  measured with forward modeling analysis of high-energy diffraction
  microscopy.
\newblock {\em Acta Materialia}, 60(10):4311--4318, 2012.

\bibitem[HP11]{hirth2011compatibility}
J.~P. Hirth and R.~C. Pond.
\newblock Compatibility and accommodation in displacive phase transformations.
\newblock {\em Progress in Materials Science}, 56(6):586--636, 2011.

\bibitem[Jam81]{james1981finite}
Richard~D. James.
\newblock Finite deformation by mechanical twinning.
\newblock {\em Archive for Rational Mechanics and Analysis}, 77(2):143--176,
  1981.

\bibitem[KF08]{kleman2008disclinations}
Maurice Kleman and Jacques Friedel.
\newblock Disclinations, dislocations, and continuous defects: A reappraisal.
\newblock {\em Reviews of Modern Physics}, 80(1):61, 2008.

\bibitem[Kha83]{khachaturian1983theory}
A.~G. Khachaturian.
\newblock {\em Theory of structural transformations in solids}.
\newblock John Wiley and Sons, New York, NY, 1983.

\bibitem[KL92]{kroner1992gauge}
E.~Kr{\"o}ner and D.~C. Lagoudas.
\newblock Gauge theory with disclinations.
\newblock {\em International Journal of Engineering Science}, 30(1):47--53,
  1992.

\bibitem[KLT06]{kinderlehrer2006variational}
David Kinderlehrer, Irene Livshits, and Shlomo Ta'Asan.
\newblock A variational approach to modeling and simulation of grain growth.
\newblock {\em SIAM Journal on Scientific Computing}, 28(5):1694--1715, 2006.

\bibitem[Kon55]{kondo1955non}
K~Kondo.
\newblock Non-{R}iemannian geometry of imperfect crystals from a macroscopic
  viewpoint.
\newblock {\em R{AAG} Memoirs of the unifying study of the basic problems in
  engineering science by means of geometry}, 1:6--17, 1955.

\bibitem[KS78]{knowles1978failure}
J.~K. Knowles and E.~Sternberg.
\newblock On the failure of ellipticity and the emergence of discontinuous
  deformation gradients in plane finite elastostatics.
\newblock {\em Journal of Elasticity}, 8(4):329--379, 1978.

\bibitem[KS79]{kleman1979tentative}
M.~Kleman and J.~. Sadoc.
\newblock A tentative description of the crystallography of amorphous solids.
\newblock {\em Journal de Physique Lettres}, 40(21):569--574, 1979.

\bibitem[KT08]{kuroda2008finite}
Mitsutoshi Kuroda and Viggo Tvergaard.
\newblock A finite deformation theory of higher-order gradient crystal
  plasticity.
\newblock {\em Journal of the Mechanics and Physics of Solids},
  56(8):2573--2584, 2008.

\bibitem[LB06]{listak2006stabilization}
Jessica Listak and Michael~R Bockstaller.
\newblock Stabilization of grain boundary morphologies in lamellar block
  copolymer/nanoparticle blends.
\newblock {\em Macromolecules}, 39(17):5820--5825, 2006.

\bibitem[Lee69]{lee}
E.~H. Lee.
\newblock Elastic-plastic deformation at finite strains.
\newblock {\em Journal of Applied Mechanics}, 36:1--6, 1969.

\bibitem[LJ12]{levitas2012advanced}
V.~I. Levitas and M.~Javanbakht.
\newblock Advanced phase-field approach to dislocation evolution.
\newblock {\em Physical Review B}, 86(14):140101, 2012.

\bibitem[LS06]{levkovitch2006large}
Vladislav Levkovitch and Bob Svendsen.
\newblock On the large-deformation-and continuum-based formulation of models
  for extended crystal plasticity.
\newblock {\em International Journal of Solids and Structures},
  43(24):7246--7267, 2006.

\bibitem[LVVZ{\etalchar{+}}02]{li2002atomistic}
Ju~Li, Krystyn~J Van~Vliet, Ting Zhu, Sidney Yip, and Subra Suresh.
\newblock Atomistic mechanisms governing elastic limit and incipient plasticity
  in crystals.
\newblock {\em Nature}, 418(6895):307--310, 2002.

\bibitem[LXBC10]{lehman2010cyclic}
L.~P. Lehman, Y.~Xing, T.~R. Bieler, and E.~J. Cotts.
\newblock Cyclic twin nucleation in tin-based solder alloys.
\newblock {\em Acta Materialia}, 58(10):3546--3556, 2010.

\bibitem[MT62]{mindlin1962effects}
R.D. Mindlin and H.F. Tiersten.
\newblock Effects of couple-stresses in linear elasticity.
\newblock {\em Archive for Rational Mechanics and Analysis}, 11(1):415--448,
  1962.

\bibitem[Mul56]{mullins1956two}
William~W. Mullins.
\newblock Two-dimensional motion of idealized grain boundaries.
\newblock {\em Journal of Applied Physics}, 27(8):900--904, 1956.

\bibitem[Nab87]{nabarro1967theory}
F.~R.~N. Nabarro.
\newblock {\em Theory of crystal dislocations}.
\newblock Dover, 1987.

\bibitem[PAN82]{peirce1982analysis}
D.~Peirce, R.~J. Asaro, and A.~Needleman.
\newblock An analysis of nonuniform and localized deformation in ductile single
  crystals.
\newblock {\em Acta Metallurgica}, 30(6):1087--1119, 1982.

\bibitem[PBSR08]{paul2008non}
Arya Paul, Jayee Bhattacharya, Surajit Sengupta, and Madan Rao.
\newblock Non-affine deformation in microstructure selection in solids ii:
  Elastoplastic theory for the dynamics of solid state transformations.
\newblock {\em Journal of Physics: Condensed Matter}, 20(36):365211, 2008.

\bibitem[RA05]{roy2005finite}
Anish Roy and Amit Acharya.
\newblock Finite element approximation of field dislocation mechanics.
\newblock {\em Journal of the Mechanics and Physics of Solids}, 53(1):143--170,
  2005.

\bibitem[RFL{\etalchar{+}}12]{ryu2012role}
Hyung~Ju Ryu, David~B Fortner, Sukbin Lee, Rachel Ferebee, Marc De~Graef,
  Konstantinos Misichronis, Apostolos Avgeropoulos, and Michael~R Bockstaller.
\newblock Role of grain boundary defects during grain coarsening of lamellar
  block copolymers.
\newblock {\em Macromolecules}, 46(1):204--215, 2012.

\bibitem[Ric76]{rice1976localization}
James~R. Rice.
\newblock The localization of plastic deformation.
\newblock In W.~T. Koiter, editor, {\em Proceedings of the 14th International
  Congress on Theoretical and Applied Mechanics, Delft}, pages 207--220.
  North-Holland Publishing Company, 1976.

\bibitem[RK09]{romanov2009application}
Alexey~E. Romanov and Anna~L. Kolesnikova.
\newblock Application of disclination concept to solid structures.
\newblock {\em Progress in Materials Science}, 54(6):740--769, 2009.

\bibitem[RLBF03]{rodney2003phase}
D.~Rodney, Y.~Le~Bouar, and A.~Finel.
\newblock Phase field methods and dislocations.
\newblock {\em Acta Materialia}, 51(1):17--30, 2003.

\bibitem[Roi78]{roitburd1978martensitic}
A.~L. Roitburd.
\newblock Martensitic transformation as a typical phase transformation in
  solids.
\newblock {\em Solid State Physics}, 33:317--390, 1978.

\bibitem[SB98]{simha1998kinetics}
N.K. Simha and K.~Bhattacharya.
\newblock Kinetics of phase boundaries with edges and junctions in a
  three-dimensional multi-phase body.
\newblock {\em Journal of the Mechanics and Physics of Solids}, 46:2323--2359,
  1998.

\bibitem[Shi73]{shield1973rotation}
R.~T. Shield.
\newblock The rotation associated with large strains.
\newblock {\em SIAM Journal on Applied Mathematics}, 25(3):483--491, 1973.

\bibitem[SKS{\etalchar{+}}10]{simon2010multiplication}
T.~Simon, A.~Kr{\"o}ger, C.~Somsen, A.~Dlouhy, and G.~Eggeler.
\newblock On the multiplication of dislocations during martensitic
  transformations in niti shape memory alloys.
\newblock {\em Acta Materialia}, 58(5):1850--1860, 2010.

\bibitem[Sle83]{slemrod1983admissibility}
Marshall Slemrod.
\newblock Admissibility criteria for propagating phase boundaries in a van der
  waals fluid.
\newblock {\em Archive for Rational Mechanics and Analysis}, 81(4):301--315,
  1983.

\bibitem[SLSB99]{shenoy1999martensitic}
S.~R. Shenoy, T.~Lookman, A.~Saxena, and A.~R. Bishop.
\newblock Martensitic textures: Multiscale consequences of elastic
  compatibility.
\newblock {\em Physical Review B}, 60(18):R12537, 1999.

\bibitem[Sok51]{sokolnikoff1951tensor}
I.~S. Sokolnikoff.
\newblock {\em Tensor analysis: {T}heory and applications}.
\newblock Wiley New York, 1951.

\bibitem[Ste96]{steinmann1996views}
P.~Steinmann.
\newblock Views on multiplicative elastoplasticity and the continuum theory of
  dislocations.
\newblock {\em International Journal of Engineering Science},
  34(15):1717--1735, 1996.

\bibitem[TCF13a]{taupin_shear_coupled}
V.~Taupin, L.~Capolungo, and C~Fressengeas.
\newblock {Disclination mediated plasticity in shear-coupled boundary
  migration}.
\newblock {\em accepted in International Journal of Plasticity}, 2013.

\bibitem[TCF{\etalchar{+}}13b]{taupin2012grain}
V.~Taupin, L.~Capolungo, C.~Fressengeas, A.~Das, and M.~Upadhyay.
\newblock Grain boundary modeling using an elasto-plastic theory of dislocation
  and disclination fields.
\newblock {\em Journal of the Mechanics and Physics of Solids}, 61:370--384,
  2013.

\bibitem[Tho34]{thomas1934systems}
T.~Y. Thomas.
\newblock Systems of total differential equations defined over simply connected
  domains.
\newblock {\em The Annals of Mathematics}, 35(4):730--734, 1934.

\bibitem[TN04]{truesdell2004non}
Clifford Truesdell and Walter Noll.
\newblock {\em The non-linear field theories of mechanics}.
\newblock Springer, 2004.

\bibitem[Tou62]{toupin1962elastic}
R.~A. Toupin.
\newblock Elastic materials with couple-stresses.
\newblock {\em Archive for Rational Mechanics and Analysis}, 11(1):385--414,
  1962.

\bibitem[Tou64]{toupin1964theories}
Richard~A. Toupin.
\newblock Theories of elasticity with couple-stress.
\newblock {\em Archive for Rational Mechanics and Analysis}, 17(2):85--112,
  1964.

\bibitem[UCTF11]{upadhyay2011grain}
M.~Upadhyay, L.~Capolungo, V.~Taupin, and C.~Fressengeas.
\newblock Grain boundary and triple junction energies in crystalline media: a
  disclination based approach.
\newblock {\em International Journal of Solids and Structures},
  48(22):3176--3193, 2011.

\bibitem[UCTF13]{upadhyay2013elastic}
Manas~Vijay Upadhyay, Laurent Capolungo, Vincent Taupin, and Claude
  Fressengeas.
\newblock Elastic constitutive laws for incompatible crystalline media: the
  contributions of dislocations, disclinations and g-disclinations.
\newblock {\em Philosophical Magazine}, 93(7):794--832, 2013.

\bibitem[VBAF06]{varadhan2006dislocation}
S.~N. Varadhan, A.~J. Beaudoin, A.~Acharya, and C.~Fressengeas.
\newblock Dislocation transport using an explicit galerkin/least-squares
  formulation.
\newblock {\em Modelling and Simulation in Materials Science and Engineering},
  14(7):1245, 2006.

\bibitem[WL10]{wang2010phase}
Yunzhi Wang and Ju~Li.
\newblock Phase field modeling of defects and deformation.
\newblock {\em Acta Materialia}, 58(4):1212--1235, 2010.

\bibitem[WSL{\etalchar{+}}13]{wang2013defective}
Y.~Morris Wang, Frederic Sansoz, Thomas LaGrange, Ryan~T. Ott, Jaime Marian,
  Troy~W. Barbee~Jr, and Alex~V. Hamza.
\newblock Defective twin boundaries in nanotwinned metals.
\newblock {\em Nature Materials}, 2013.

\bibitem[ZCA13]{zhu-chapman}
Y.~Zhu, S.~J. Chapman, and A.~Acharya.
\newblock {Dislocation motion and instability}.
\newblock {\em in press Journal of the Mechanics and Physics of Solids}, 2013.

\bibitem[ZLJVV{\etalchar{+}}04]{zhu2004predictive}
Ting Zhu, Ju~Li, Krystyn J~Van~Vliet, Shigenobu Ogata, Sidney Yip, and Subra
  Suresh.
\newblock Predictive modeling of nanoindentation-induced homogeneous
  dislocation nucleation in copper.
\newblock {\em Journal of the Mechanics and Physics of Solids}, 52(3):691--724,
  2004.

\bibitem[ZTY{\etalchar{+}}10]{zarnetta2010identification}
Robert Zarnetta, Ryota Takahashi, Marcus~L Young, Alan Savan, Yasubumi Furuya,
  Sigurd Thienhaus, Burkhard Maa{\ss}, Mustafa Rahim, Jan Frenzel, Hayo
  Brunken, Yong~S. Chu, Vijay Srivastava, Richard~D. James, Ichiro Takeuchi,
  Gunther Eggeler, and Alfred Ludwig.
\newblock Identification of quaternary shape memory alloys with near-zero
  thermal hysteresis and unprecedented functional stability.
\newblock {\em Advanced Functional Materials}, 20(12):1917--1923, 2010.

\end{thebibliography}
\bibliographystyle{alpha}

\end{document}